\documentclass[12pt,preprint]{aastex}

\newcommand{\pks}{PKS~2155--304}
\newcommand{\xs}{$\sigma_{\rm XS}^{2}$}
\newcommand{\sqxs}{$\sigma_{\rm XS}$}
\newcommand{\nxs}{$\sigma_{\rm NXS}^{2}$}
\newcommand{\fvar}{$F_{\rm var}$}
\newcommand{\fpp}{$F_{\rm pp}$}
\newcommand{\bhm}{{M$_{\rm BH}$}}
\newcommand{\nbf}{{$\nu_{\rm bf}$}}

\newcommand{\cha}{{\it Chandra}}
\newcommand{\xmm}{{\it XMM-Newton}}
\newcommand{\sax}{{\it BeppoSAX}}
\newcommand{\asca}{{\it ASCA}}
\newcommand{\xte}{{\it RXTE}}

\newcommand{\et}{et al.\ }
\newcommand{\beq}{\begin{equation}}
\newcommand{\eeq}{\end{equation}}

\slugcomment{Accepted by The Astrophysical Journal}

\shortauthors{Zhang et al. }
\shorttitle{X-ray Variability of PKS~2155--304 with \xmm}

\begin{document}

\title{XMM-Newton View of PKS~2155--304: Characterizing the X-ray Variability Properties with EPIC-PN}

\authoraddr{ }

\author{Y.H. Zhang}
\affil{Department of Physics and Tsinghua Center for Astrophysics (THCA), Tsinghua University, Beijing 100084, P.R. China}
\email{youhong.zhang@mail.tsinghua.edu.cn}
\author{A. Treves}
\affil{Dipartimento di Scienze, Universit\`a degli Studi dell'Insubria,
         via Valleggio 11, I-22100 Como, Italy}
\author{A. Celotti}
\affil{International School for Advanced Studies, SISSA/ISAS,
         via Beirut 2-4, I-34014 Trieste, Italy}
\author{Y.P. Qin and J.M. Bai}
\affil  {National Astronomical Observatories/Yunnan Observatory, Chinese Academy of Sciences, P. O. Box 110, Kunming, Yunnan, 650011, P. R. China}

\begin{abstract}

Starting from \xmm\ EPIC-PN data, we present the X-ray variability characteristics of \pks\ using a simple analysis of the excess variance, \xs, and of the fractional rms variability amplitude, \fvar. The scatter in \xs\ and \fvar, calculated using 500~s long segments of the light curves, is smaller than the scatter expected for red noise variability. This alone does not imply that the underlying process responsible for the variability of the source is stationary, since the real changes of the individual variance estimates are possibly smaller than the large scatters expected for a red noise process. In fact the averaged \xs\ and \fvar, reducing the fluctuations of the individual variances, change with time, indicating non-stationary variability. Moreover, both the averaged \sqxs\ (absolute rms variability amplitude) and \fvar\ show linear correlation with source flux but in an opposite sense: \sqxs\ correlates with flux, but \fvar\ anti-correlates with flux. These correlations suggest that the variability process of the source is strongly non-stationary as random scatters of variances should not yield any correlation. \fvar\ spectra were constructed to compare variability amplitudes in different energy bands. We found that the fractional rms variability amplitude of the source, when significant variability is observed, increases logarithmically with the photon energy, indicating significant spectral variability. The point-to-point variability amplitude may also track this trend, suggesting that the slopes of the power spectral density of the source are energy-independent. Using the normalized excess variance the black hole mass of \pks\ was estimated to be about $1.45 \times 10^8 M_{\bigodot}$. This is compared and contrasted with the estimates derived from measurements of the host galaxies. 

\end{abstract}



\section{Introduction}\label{sec:intro}

Blazars are the extreme subclass of Active Galactic Nuclei (AGNs), they show rapid variability on different timescales across the whole electromagnetic spectrum (Ulrich, Maraschi \& Urry 1997). A number of observational facts define our current understanding of blazars: the entire electromagnetic emission from blazars is produced by relativistic electrons tangled with the magnetic field in a relativistic jet roughly aligned with our line of sight (Blandford \& Rees 1978; see Urry \& Padovani 1995 for a review). TeV blazars represent a subclass whose emission has been detected up to TeV energies with ground-based Cerenkov telescopes. The number of TeV blazars is gradually growing. Up to now, there are 6 confirmed TeV blazars, including the three prototypical blazars Mrk~421, Mrk~501, and PKS~2155$-$304. The overall spectral energy distributions of TeV blazars show that synchrotron emission from these sources peaks at the high energy (UV/soft X-ray) band. This indicates that the X-ray emission from TeV sources is the high energy tail of the synchrotron emission component produced in the inner part of the relativistic jets,  where the most rapid variability is expected. Therefore, TeV blazars are jet/synchrotron emission dominated X-ray sources. They have been the important targets of various X-ray telescopes such as \asca, \sax, \xte, \cha, and \xmm. These X-ray observations have revealed very complex variability patterns for the TeV sources (for a review see Pian 2002).

X-ray emission from TeV blazars is known to be the most variable. The X-ray light curves obtained so far with various X-ray telescopes show aperiodic and unpredictable events (e.g., flare intensity and duration), though the flares usually occur on timescales of day as viewed from the long-look observations (e.g., Tanihata \et 2001; Zhang \et 2002; Cui 2004; Massaro \et 2004). Clearly characterizing the X-ray variability of TeV blazars is important for exploring the underlying physical process at work. However, the X-ray variability of TeV blazars is a stochastic red noise process, and its analysis requires large amount of data in order to obtain  meaningful {\it average} properties of the statistical moments of the light curves. Probably the best tool to characterize the X-ray variability is to measure the fluctuation Power Spectral Density (PSD). The PSD method has been frequently used for examining the X-ray variability properties of X-ray binaries and Seyfert galaxies (e.g., Pottschmidt 2003; Markowitz \et 2003). The source's PSD represents the amount of mean variability amplitude as a function of temporal frequency (timescale$^{-1}$). Measuring the PSD requires rather long, high-quantity (evenly sampled) and adequate data (multiple light curves or multiple segments of a long observation)  for the average PSD to have physical significance. The PSDs obtained so far for TeV blazars are only in the high frequency range, roughly between $10^{-5}$ and $10^{-3}$~Hz. In this  range the PSDs are most likely represented by a power-law ($\mathcal{P}(f) \propto f^{-\alpha}$, where $\mathcal{P}(f)$ is the power at frequency $f$) with slope $\alpha \sim 2-3$ (Kataoka \et 2001; Zhang \et 1999, 2002; Zhang 2002; Brinkmann \et 2003). However, because of periodic gaps associated with the observations gathered with low-Earth orbital satellites such as \asca , \sax\ and \xte , usually one can not perform a full PSD analysis with such data. Special treatments have to be imposed on the data before calculating a PSD 
\footnote{PSD measurement ideally requires evenly-sampled light curves. For a light curve observed with periodic (orbital) gaps, even sampling can be obtained by interpolating gaps, if the gaps comprise only a small fraction of the total light curve, or binning on either of two timescales: binning on short timescale, say 256~s, for the duration of one single orbital light curve, and binning on orbital timescale for the duration of the monitoring (e.g., Zhang \et 1999; 2002)}. Instead, a structure function (SF) that works with light curves having gaps, has been used to perform a similar analysis but in the time domain. Physically the PSD and SF are identical only in the limit of a light curve length with $T \rightarrow \infty$ and binsize $\Delta t \rightarrow 0$ (Paltani 1999), so the characteristic quantities (break timescale and slope) derived from these two methods usually are inconsistent with each other (Kataoka \et 2001; Zhang \et 2002).

In blazar timing studies, the variability can also be characterized in terms of excess variance. The variance estimate provides a simple and straightforward means of quantifying the X-ray variability. However, there is a  rather large scatter in the variance associated with the stochastic nature of red noise variability. This implies that only the mean variance averaged from a large number of data sets is meaningful. Previous variance estimates of blazars were usually calculated using only a single light curve (e.g., Zhang \et 2002; Ravasio \et 2004). Quantitative comparisons of individual variances estimated in such a way, usually referring to the repeated observations of the same source at different epochs, can be misleading, because an observed light curve is only a single realization of a stochastic red noise process (Press 1978). Such a comparison is desirable since one wants to know if the underlying process responsible for the variability evolves with time, i.e., the stationarity of the variability. Only real changes in the variance (or PSD), quantified with the mean variance (or PSD) averaged from a number of light curves or a number of segments of a long light curve, would reflect changes in the physical conditions of the variability process.

Recently, Vaughan \et (2003b) explored some practical aspects of measuring the amplitude of variability in ``red noise'' light curves typical of  AGNs. They examined the statistical properties of the quantities commonly used to estimate the variability amplitude in AGN light curves, such as excess variance, \xs, and the fractional rms variability amplitude, \fvar. Using a long, consecutive \xmm\ light curve, they explored the variability properties of a bright Seyfert 1 galaxy Markarian 766. The source is found to show a linear correlation between absolute rms variability amplitude and flux, and to show significant spectral variability.

In this paper we will use variance estimators to examine the variability properties of all public \xmm\ observations for \pks\ taken over a period of about three years (12 exposures from 6 orbits). The EPIC-PN data are most suitable for such an analysis because of their high count rates. The observations and data reduction process are presented in \S~\ref{sec:obs}. \S~\ref{sec:review} discusses some practical aspects of the variance estimator. The results are presented in \S~\ref{sec:results}, and discussed in \S~\ref{sec:disc}. The conclusions are summarized in \S~\ref{sec:conc}.


\section{The XMM-Newton Observations}\label{sec:obs}

\pks\ was observed during six orbits of \xmm\ for about 100~ks each over a period of about 3.5 years. The source was the target of instrument calibration during 4 out of the six orbits. In this paper, we will concentrate on data obtained with the PN camera that are less affected by photon pile-up and have better time resolution. There are a total of 12 PN exposures, each lasting about 40~ks. The PN camera was operated in Small Window (SW) mode during 10 exposures, and in Timing mode during 2 exposures. Different filters (thin, medium and thick) were used. The properties of each exposure are detailed in Table~\ref{tab:obs}, where an identification number is allocated to each exposure.
 
All PN data were reprocessed using the \xmm\ Science Analysis System (SAS) 6.0 and the latest available calibration data.  First we checked the high particle background periods caused by solar activity. We computed the hard ($E>10$~keV) count rate and discarded the time intervals where the count rate is significantly higher and variable. We then checked the photon pile-up for imaging mode, which is strong for a bright source such as \pks . We determined the central region to be discarded by using the SAS task {\it epatplot} on different circular and annular regions of each image. The source data were extracted from rings centered on the source position. The inner radius was determined by photon pile effects, usually the radius of the central circular region to be discarded is 10''. The outer radius, ranging from 35'' to 40'', was fixed by the position of the source on the chip. The timing mode was not affected by the photon pile up effect, and we extracted the source photons from 10 RAWX pixels wide region centered on the brightest strip of the source. In order to minimize the photon pile up effect only single pixel events (pattern=0) with quality flag=0 were selected. The background events were extracted from regions least effected by source photons. 

In summary, in imaging mode, exposure 545-1 and 724 were not significantly affected by photon pile up effects, while other exposures in imaging mode were influenced by this effect. High particle background occurred in the middle and in the end of exposure 174-2, and in the end of exposure 545-2 and 724. In our analysis, we discard exposure 87-1 and 87-2 because the high particle background occurred throughout the whole exposure. Moreover, we also drop the exposure 545-2 due to a calibration problem in timing mode. Detailed information about high particle background and photon pile effects is shown in Table~\ref{tab:obs}. The observation during orbit 174 was preliminarily analyzed in Maraschi \et (2004) and Zhang \et (2004).

         
\section{Random Process and Variance Estimators}\label{sec:review}

Before performing data analysis on the \xmm\ light curves mentioned in the last section, it is useful to review some specific aspects related to a variance analysis of the X-ray variability.

\subsection{Integrated PSD and Variance}\label{sec:psdvar}

Parseval's theorem (see, e.g., van der Klis 1989; Press \et 1992) shows that the integral of the PSD between two Fourier frequencies $f_{1}$ and $f_{2}$ ($f_1 < f_2$) yields the expectation value of the `true' variance due to variations between the corresponding timescales ($1/f_{2}$ and $1/f_{1}$)

\begin{equation}
\label{eqn:correspondence1}
\langle S^{2} \rangle = \int_{f_{1}}^{f_{2}} \mathcal{P}(f) df.
\label{eq:varpsd}
\end{equation}

An astronomical light curve, usually a discrete time series $x_i$, is just a realization of a random process, and the integrated periodogram gives rise to the observed variance for that particular realization

\begin{equation}
\label{eqn:correspondence}
S^{2} 
= \sum_{j=1}^{N/2} P(f_{j}) \Delta f,
\end{equation}
where $\Delta f$ is the frequency resolution of the Discrete Fourier Transform (DFT) for the discrete time series ($\Delta f=1/(N\Delta T)$, $\Delta T$ is the binsize of the light curve). The total variance of a real light curve is equal to its periodogram integrated from the frequency range
$f_{1}=1/(N\Delta T)$ to $f_{\rm Nyq}=1/(2\Delta T)$ (Nyquist frequency). It is important to note that the periodogram $P(f_{j})$ is only one realization of the underlying $\mathcal{P}(f)$.

The observed variance can be simply obtained from the light curve:  
\begin{equation}
\label{eqn:variance}
S^{2} = \frac{1}{N-1} \sum_{i=1}^{N} (x_{i} - \bar{x})^{2},
\end{equation} 
where $\bar{x}$ is the arithmetic mean of $x_{i}$. This variance is usually different from observation to observation (see discussion below). In the limit of large $N$ the variance estimate from the light curve should be identical to the one integrated from the corresponding periodogram. 

Equation (\ref{eqn:variance}) also indicates that variance is proportional to the square of the count rate. For example, if the count rate obtained from a detector (e.g., \xmm\ PN) is a factor m larger than the one simultaneously obtained from another detector (e.g., \xmm\ MOS1 or MOS2 ), the variance obtained from the former detector will be $m^2$ factor larger than the one simultaneously obtained with the latter detector. However, this does not imply the source variability is different from the two detectors at the same time. Therefore, in order to compare the variance obtained from different detectors, or  from the observations at different epochs, or between different sources, the normalized variance, $S^{2}/\bar{x}^{2}$, is usually used.  

\subsection{Excess variance and $F_{\rm var}$}\label{sec:fvar}

A real light curve $x_i$ has finite uncertainties $\sigma_{i}$ due to measurement errors (Poisson noise for an X-ray photon counting signal). The uncertainties due to photon counting also yield an additional variance that should be subtracted in order to obtain the {\it intrinsic} variance. This is the so-called `excess variance' (Nandra \et 1997; Edelson \et 2002) 

\begin{equation}
\label{eqn:excess_variance}
\sigma_{\rm{XS}}^{2} = S^{2} - \overline{\sigma^2},
\end{equation}
where $\overline{\sigma^2}$ is the mean error squared

\begin{equation}
\label{eqn:mean_error}
\overline{\sigma^{2}} = \frac{1}{N}\sum_{i=1}^{N} \sigma_{i}^{2}.
\end{equation}
\xs\ is an absolute quantity that linearly correlates with count rate. Because the count rates are independent, the normalized excess variance, simply given by $\sigma_{\rm{NXS}}^{2}=\sigma_{\rm{XS}}^{2}/\bar{x}^{2}$, is often used for comparing variances between different observations (or different segments of a long observation) of the same source, and between different sources. The square root of \nxs\ is the fractional root mean square (rms) variability amplitude, \fvar ,  a common measure of the intrinsic variability amplitude that corrects the effects of the measurement errors (Edelson, Pike \& Krolik 1990; Rodr\'{\i}guez-Pascual \et 1997)
\begin{equation}
\label{eqn:fvar}
F_{\rm{var}} = \sqrt{ \frac{S^{2} -
\overline{\sigma^{2}}}{\bar{x}^{2}}}.
\end{equation} 


\subsection{Intrinsic scatter in variance}
\label{sec:scatter}

Since red noise variability is a stochastic process, one should expect random fluctuations in both the mean and variance with time (between segments of a long observation or between observations taken at different epochs). For a stationary process whose statistical properties do not evolve with time, the distribution of the individual variances from red noise with a steep PSD has a non-Gaussian shape with rather large scatter (Vaughan \et 2003b). This scatter is intrinsic to the stationary process, which is caused by the limited length of a given observation. This means that the red noise variability is weakly non-stationary (Press \& Rybicki 1997). Thus the scatter in the variance is not caused either by the measurement errors or by the number of data points used. However, the underlying process responsible for the variability itself may also change with time (e.g., if the PSD and variance changes with time). If this is the case, the variability is said to be strongly non-stationary. Real changes in the variability process could provide insight into the changing physical conditions in the nuclear emission region, while the random changes of statistical moments expected for a stationary red noise process yield no physical insight. As the purpose of timing series analysis is to gain insight into the physical process, it is desirable to discriminate the changes in variance produced by a time-stationary process and by a strong non-stationary process. Therefore, all the subsequent non-stationarity discussed hereafter throughout the text refers to strong non-stationarity.

The stochastic nature of a red noise process implies that only the mean properties of the statistical quantities averaged from a number of data sets can provide physical insight. Averaging over many independent realizations reduces the random fluctuations inherent in the red noise process. The expectation values, not one realization, of the statistical quantities (such as variance) should be representative of the properties of the underlying process. In order to search for non-stationary variability, an ensemble of light curves (or of short segments of a long light curve) is therefore needed. For example, comparing different PSD is the method frequently employed in timing analysis of X-ray binaries. The mean PSDs averaged from the periodograms of a large number of light curves show significant differences from epoch to epoch, indicating that the variability process of X-ray binaries is strongly non-stationary (e.g., van der Klis 1995). Moreover, the way in which the variability properties evolve with time, is even more important, from which one can effectively infer the detailed working mechanisms in these systems (e.g., Belloni \& Hasinger 1990; Uttley \& M$^{\rm c}$Hardy 2001; Belloni, Psaltis \& van der Klis 2002; Pottschmidt \et 2003). 

However, unlike X-ray binaries, AGN data are usually not adequate for performing a full PSD analysis. In order to search for non-stationarity of the variability in a number of short light curves (or of short segments of a long light curve), one can test whether the variances differ significantly. Vaughan \et (2003b) proposed three practical methods as variance estimators: (1) the individual variance estimates are compared with the expected scatter around the mean. Here, the expected scatter is calculated using Monte Carlo simulations of a stationary processes. Because the intrinsic scatter in the variance of a stationary process is rather large for red noise data, this method is only sensitive to very large changes in the variability amplitude. Moreover, one has to assume a PSD shape for the analyzed source; (2) the variances are averaged at various epochs by binning a number ($N \ge 20$) of individual variance estimates. Here, the light curve segments from which the individual variance are calculated, should have the same sampling properties in binsize, length and data points. This is most useful when searching for subtle changes in variability amplitude but requires large datasets (in order that the variance can be sufficiently averaged); (3) individual variance estimates can be sorted and binned by count rates so that relationship between variance or \fvar\ and count rate can be constructed. The existence of a correlation would imply that the variability process is strongly non-stationary, as the random fluctuations in variance should not produce such a correlation. A linear correlation between binned, absolute rms variability amplitude and flux has been found in X-ray binaries and Seyfert galaxies, indicating non-stationarity of the X-ray variability (Uttley \& M$^{\rm c}$Hardy 2001).


\section{Results}\label{sec:results}

\subsection{Changes of the \xmm\ Light Curves}\label{sec:stationarity}

\begin{figure}
\epsscale{0.51}
\plotone{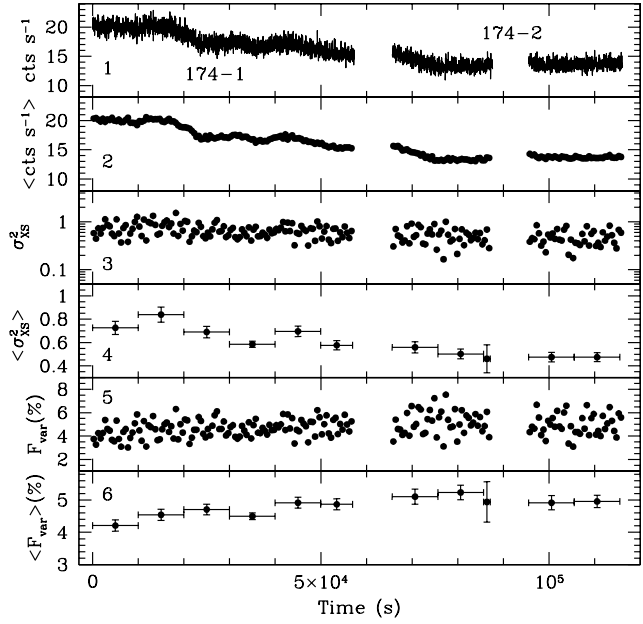}
\caption { \footnotesize Panel 1: background subtracted, 25~s binned 0.2--10~keV PN light curve of PKS~2155--304 obtained in Revolution 174. Note that there is small offset of count rate between 174-1 and 174-2 due to different extraction region of the source photons. High particle background occurred in the middle and in the end of 174-2 part. Panel 2 and 3: mean count rate and excess variance calculated from 20 consecutive points of panel 1. Panel 2 corresponds to the light curve binned over 500~s. Panel 4: mean excess variance averaged by binning 20 continuous individual excess variances. This average excess variance is inconsistent with constant. Panel 5: fractional rms variability amplitude calculated from 20 continuous points of panel 1. Panel 6: mean fractional rms variability amplitude averaged by binning 20 continuous individual amplitudes. This average fractional rms variability amplitude is inconsistent with constant, either. }
\label{fig:lc:174}
\end{figure}

\begin{figure}
\plotone{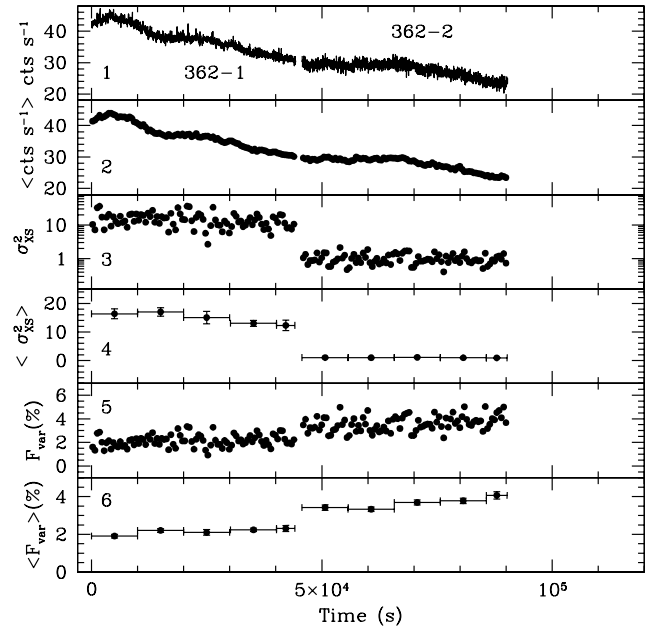}
\caption { \footnotesize Same as Figure~\ref{fig:lc:174} but using light curve obtained during revolution 362. Panel 1: the count rates of 362-1 (timing mode) were scaled down by a factor of 4.7 in order to smoothly connect with 362-2 (imaging mode). Panel 4: the average excess variance of 362-1 is inconsistent with constant. Panel 6: the average fractional rms variability amplitude is inconsistent with constant, either.}
\label{fig:lc:362}
\end{figure}

\begin{figure}
\plotone{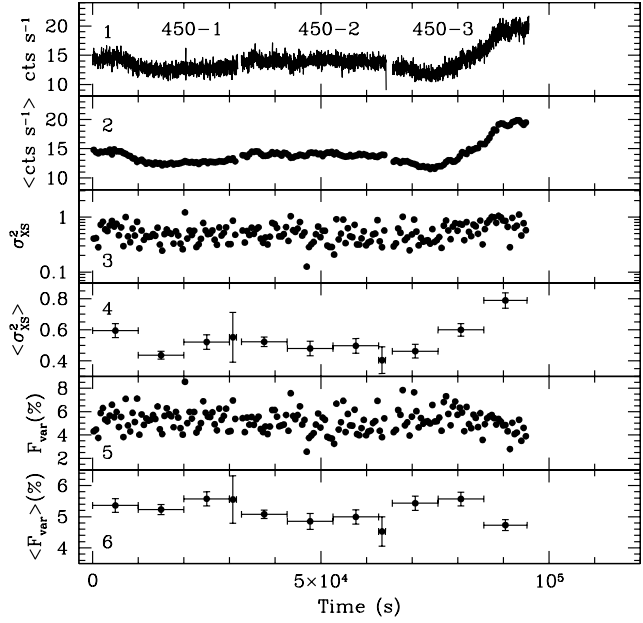}
\caption { \footnotesize Same as Figure~\ref{fig:lc:174} but using light curve obtained during revolution 450. Due to different filters and extraction regions used for each exposure there are small offsets of count rates between different exposures. Panel 4: the average excess variance of 450-1 and 450-3 is inconsistent with constant. Panel 6: the average fractional rms variability amplitude of 450-3 is inconsistent with constant, either.}
\label{fig:lc:450}
\end{figure}

\begin{figure}
\plotone{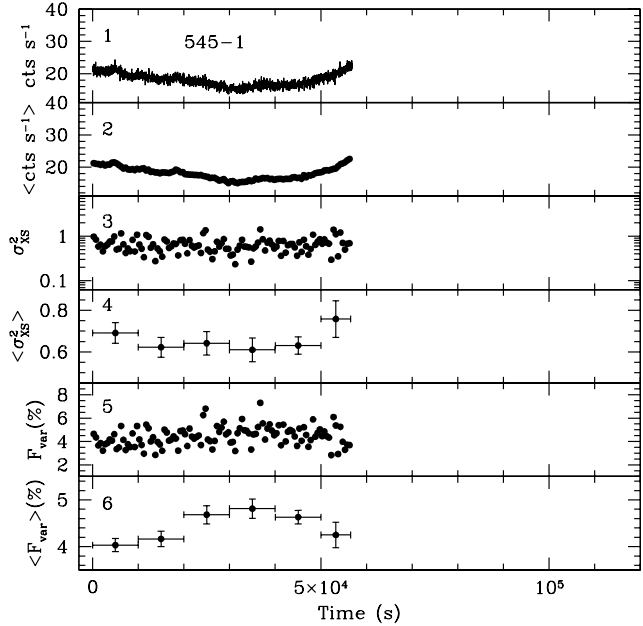}
\caption { \footnotesize Same as Figure~\ref{fig:lc:174} but using light curve obtained during revolution 545. Only 545-1 data are shown, 545-2 data are not shown due to calibration problem. Panel 6: the average fractional rms variability amplitude is inconsistent with constant. }
\label{fig:lc:545}
\end{figure}

\begin{figure}
\plotone{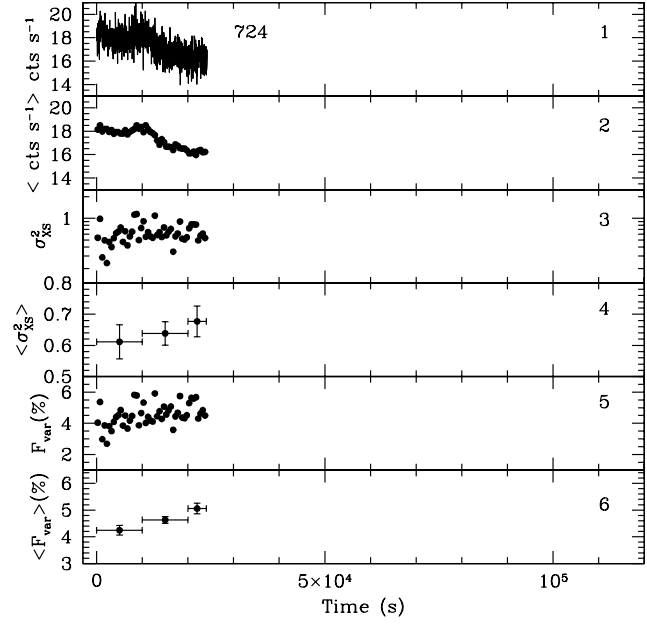}
\caption { \footnotesize Same as Figure~\ref{fig:lc:174} but using light curve obtained during revolution 724. High particle background occurred in the end of this observation. Panel 6: the average fractional rms variability amplitude is inconsistent with constant. Due to short observation length, we do not discuss it in the text.}
\label{fig:lc:724}
\end{figure}

The background subtracted, 25~s binned, broad band 0.2--10~keV light curves extracted from the PN camera are shown in Figures~\ref{fig:lc:174}--\ref{fig:lc:724} (panel 1). The length of the observation for a full revolution (i.e., Rev. 174, 362, 450, and 545) is about $10^5$~s. The source showed significant variability during each revolution, but, unfortunately, no complete flares were obtained due to the interruption of the observations. It is likely that a pronounced flare occurred at the beginning of Rev. 362 and at the end of Rev. 450. The observations also showed that the variability of \pks\ (and Mrk~421, Brinkmann \et 2003; Ravasio \et 2004) is dominated by somewhat smooth flares rather than by rapid ``flickers'' in the X-ray light curves typical of Seyfert 1 galaxies, e.g., \xmm\ light curves of MCG-6-30-15 (Vaughan, Fabian \& Nandra 2003a) and Mrk~766 (Vaughan \et 2003b). Such qualitative differences in the X-ray light curves may reflect the different mechanisms producing the variability in the two classes, i.e., TeV blazars are jet-dominated systems while Seyfert 1 galaxies are disk-dominated systems. It further indicates that the PSD slopes of TeV blazars might be steeper than those of Seyfert 1 galaxies.

\begin{figure}
\plotone{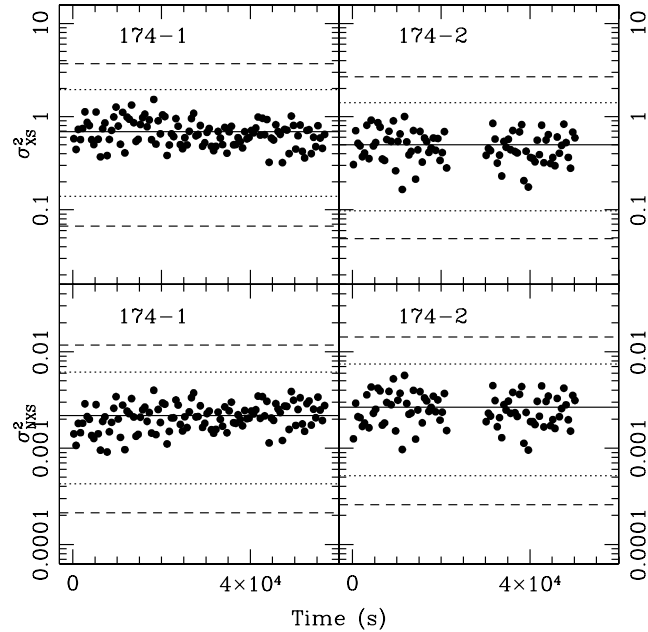}
\caption { \footnotesize Excess variance (top panel) of PKS~2155--304 same as shown in panel 3 of Figure~\ref{fig:lc:174} and the corresponding normalized excess variance (bottom panel). The solid line marks the mean variance, and the dotted and dashed line the 90\% and 99\% confidence intervals around the mean, respectively. The variances fall within the limits as expected for a stationary process. }
\label{fig:var:174}
\end{figure}

\begin{figure}
\plotone{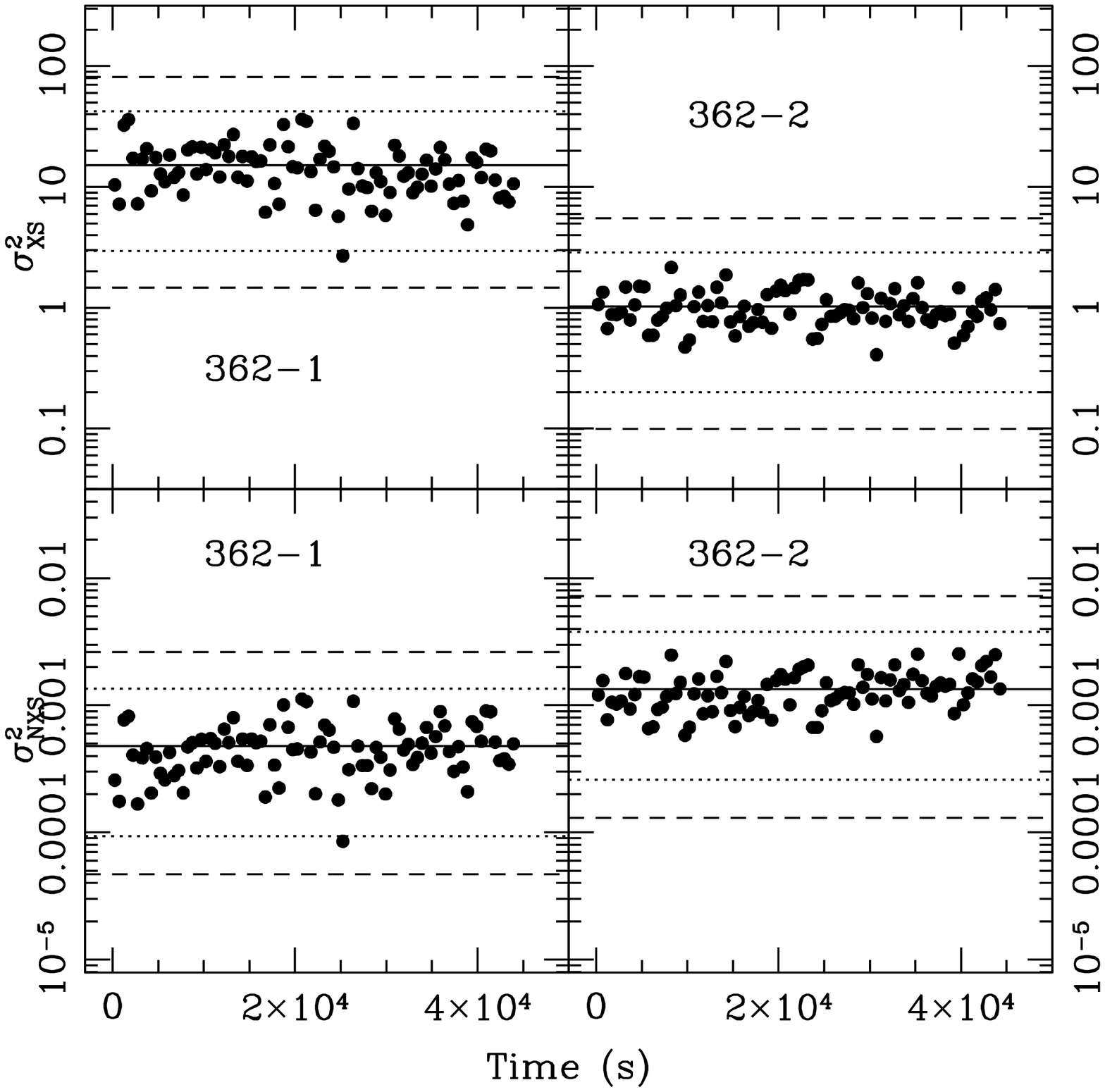}
\caption { \footnotesize Same as Figure~\ref{fig:var:174} but using revolution 362 data. The variances fall within the limits as expected for a stationary process. }
\label{fig:var:362}
\end{figure}

\begin{figure}
\plotone{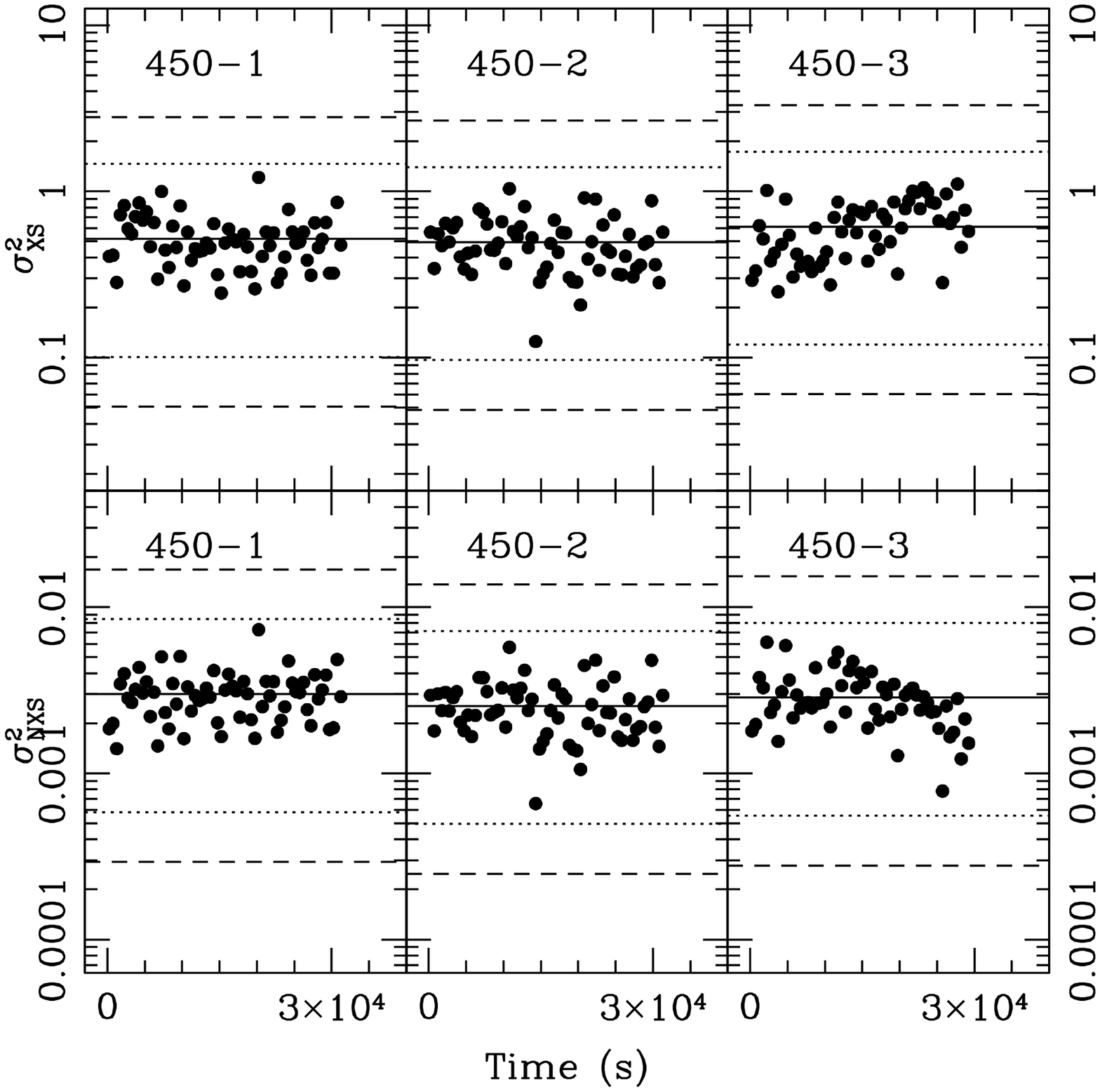}
\caption { \footnotesize Same as Figure~\ref{fig:var:174} but using revolution 450 data. The variances fall within the limits as expected for a stationary process. }
\label{fig:var:450}
\end{figure}

\begin{figure}
\plotone{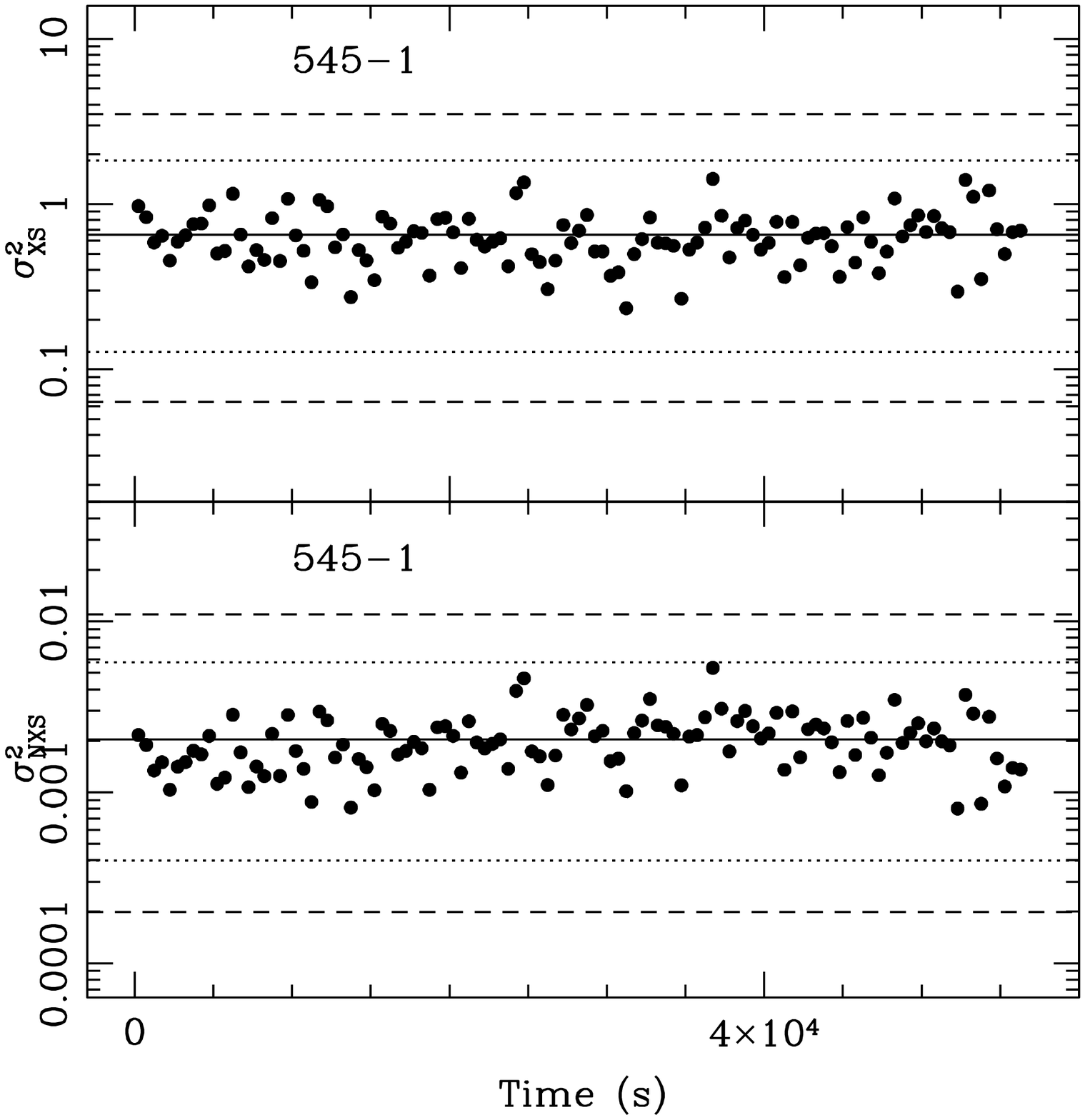}
\caption { \footnotesize Same as Figure~\ref{fig:var:174} but using revolution 545-1 data. The variances fall within the limits as expected for a stationary process. }
\label{fig:var:545}
\end{figure}

In order to test the stationarity of the light curves, we calculated the mean count rate and excess variance for every 20 consecutive data points in the light curves. This corresponds to ``instantaneous'' estimates of the source variance over a timescale of 50--500~s. The results are shown in panels 2 and 3 of Figures~\ref{fig:lc:174}--\ref{fig:lc:724}. Panel 2 corresponds to the light curve binned over a timescale of 500~s. As expected for red noise variability, the individual source variances (panel 3) show large changes within each revolution. The expected range for the excess variance and normalized excess variance, calculated using the simulation results (Table~1 of Vaughan et al. 2003b) under the assumption of a PSD slope  $\alpha = 2.0$, is plotted in Figures~\ref{fig:var:174}--\ref{fig:var:545}. Normalized excess variance, corresponding to the individual \fvar\ estimates in panel 5 of Figures~\ref{fig:lc:174}--\ref{fig:lc:724}, can be compared for different exposures of each revolution (and of different revolutions). Neither of these figures show fluctuations of source variance larger than the range expected for a stationary red noise process. Note that the confidence ranges marked here should be considered as approximate values since the PSD slope of \pks\ is probably larger than 2 (e.g., Zhang \et 2002), yielding larger confidence ranges. These tests are consistent with the variability of \pks\ being a stationary process. As stated before, however, the real changes of source variances may be masked by the rather large scatter in variance expected for stationary red noise variability.

Subtle changes in the variability amplitude can be identified by calculating the mean variance and its error through binning the consecutive individual variance estimates. The \xmm\ data of \pks\ are sufficient to examine changes of the average variance with time, allowing a more sensitive test for non-stationarity of the variability. We calculated the mean excess variance by binning 20 consecutive individual excess variance estimates, and assigning an error bar using equation (4.14) of Bevington \& Robinson (1992). The results are shown in Panel 4 of  Figures~\ref{fig:lc:174}--\ref{fig:lc:724}. Significant changes in the average variance with time are revealed for exposures 174-1, 362-1, 450-1 and 450-3, when the source showed significant variations. In order to test the significance of such changes, we fitted the mean variance with a constant for each exposure. The fitting can be rejected at larger than 95\% confidence for exposures 174-1 ($\chi^2 = 21.2$ for 5 degrees of freedom ({\it dof})), 450-1 ($\chi^2 = 10.2$ for 3 {\it dof}) and 450-3 ($\chi^2 = 24.2$ for 2 {\it dof}), and at 90\% confidence for exposure 362-1 ($\chi^2 = 7.6$ for 4 {\it dof}), respectively, indicating the mean excess variance is not consistent with a constant hypothesis, i.e., the average excess variance changes with time. For other exposures, the changes in mean variance are consistent with a constant hypothesis due to small variations or short exposure length. The results indicate that the variability of \pks\ tends to be non-stationary during flare states.

To compare variability amplitude in different exposures, we further calculated \fvar\ and its average in the same way as the excess variance. Panels 5 and 6 of Figures~\ref{fig:lc:174}--\ref{fig:lc:724} show individual \fvar\ estimates and their averages as a function of time. Each point of \fvar\ and its average corresponds to that of each \xs\ and its average. The fitting with a constant showed that the mean \fvar\ is inconsistent with a constant at larger than 95\% confidence for exposures 174-1, 362-1, 362-2, 450-3, 545-1 and 724. 

Therefore, our analysis shows that the variability of \pks\ does show genuine non-stationarity in the sense of both the absolute and the fractional variability amplitude (i.e., \xs\ and \fvar ).

\subsection{Flux-rms Correlations}\label{sec:correlation}

Figures~\ref{fig:lc:174}--\ref{fig:lc:724} (panel 4 and 6) show that the mean \xs\ and \fvar\ have a tendency to track the mean count rate but in an opposite sense: \xs\ tends to increase while \fvar\ to decrease with increasing count rate. This trend is difficult to discern from the individual \xs\ and \fvar\ estimates (panel 3 and 5 of Figures~\ref{fig:lc:174}--\ref{fig:lc:724}) due to large scatter. 

\begin{figure}
\plotone{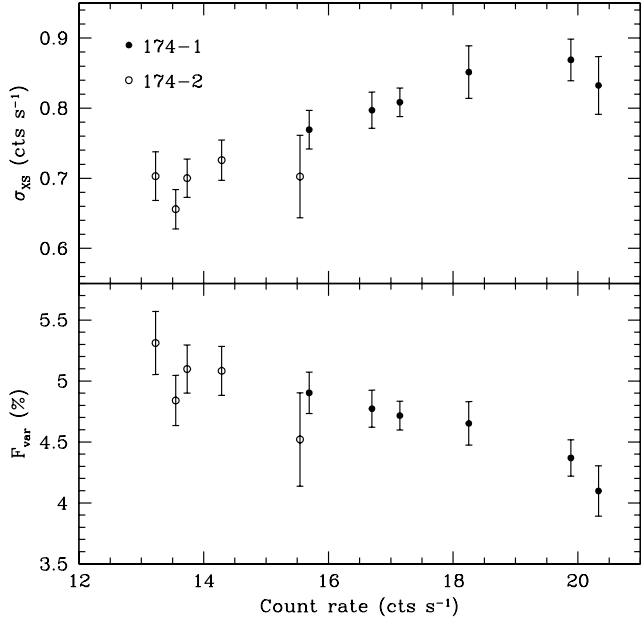}
\caption { \footnotesize The mean absolute rms variability amplitude (\sqxs $=\sqrt{\sigma_{\rm XS}^2}$, top panel) and the mean fractional rms variability amplitude ($F_{\rm var}$, bottom panel) as a function of count rate. Clearly the absolute rms amplitude correlates with flux, but the fractional rms amplitude anti-correlates with flux. The small offset of count rates between different exposures do not alter such correlation. }
\label{fig:cor:174}
\end{figure}

\begin{figure}
\plotone{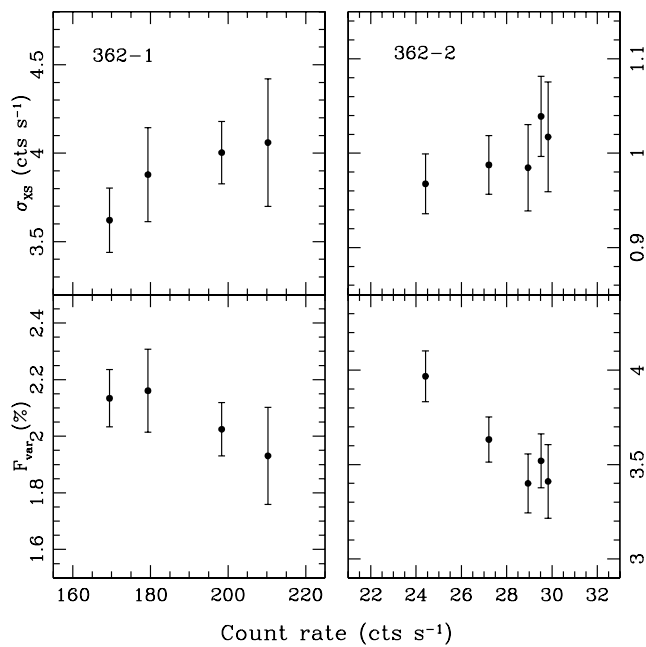}
\caption { \footnotesize Same as Figure~\ref{fig:cor:174} but using revolution 362 data. For both exposures, the absolute rms variability amplitude correlates with flux, but the fractional rms variability amplitude anti-correlates with flux. }
\label{fig:cor:362}
\end{figure}

\begin{figure}
\plotone{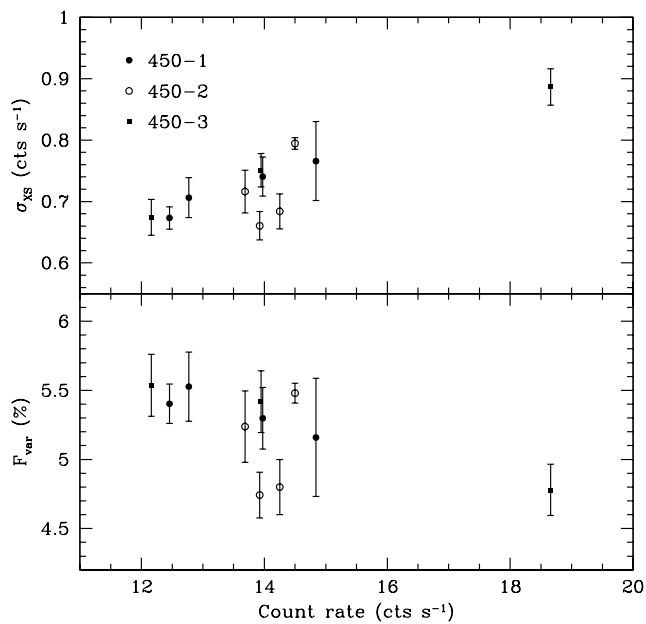}
\caption { \footnotesize Same as Figure~\ref{fig:cor:174} but using revolution 450 data. Clearly the absolute rms variability amplitude correlates with flux, but the fractional rms variability amplitude anti-correlates with flux. The small offset of count rates between different exposures do not alter such correlation.}
\label{fig:cor:450}
\end{figure}

\begin{figure}
\plotone{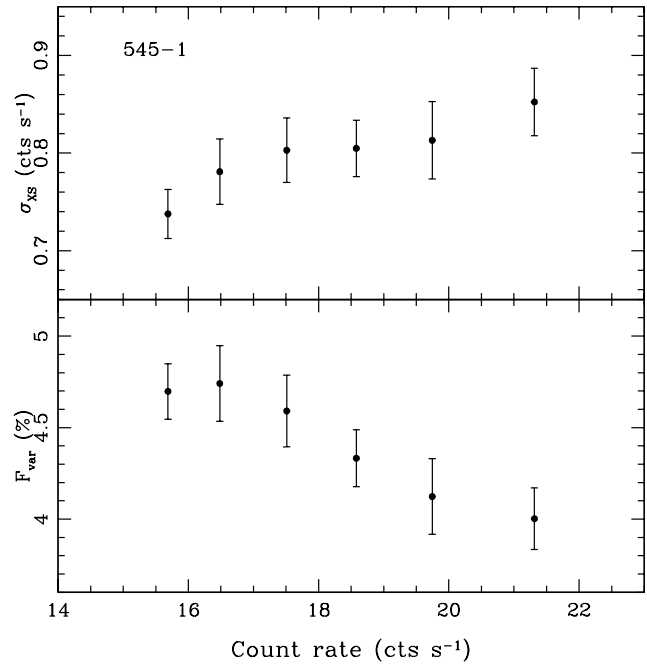}
\caption { \footnotesize Same as Figure~\ref{fig:cor:174} but using revolution 545-1 data. Clearly the absolute rms variability amplitude correlates with flux, but the fractional rms variability amplitude anti-correlates with flux. }
\label{fig:cor:545}
\end{figure}

In order to quantify the dependence of excess variance on the count rate, we sorted individual excess variance estimates by count rate and binned \xs\ every 20 estimates, the error on the mean \xs\ was calculated in the standard way as above. The results are shown in Figures~\ref{fig:cor:174}--\ref{fig:cor:545} (top panel), where the mean absolute rms amplitude (i.e., the square root of the excess variance, $\sqrt{\sigma_{\rm XS}^{2}}$) is shown as a function of mean count rate. These figures clearly show that the absolute variability amplitude correlates linearly with the flux. The offsets of count rates for the different exposures of Revolution 174 and 450 are negligible, so we put them in a single plot. The source therefore does show strong non-stationarity: the absolute variability amplitude increases, on average, as the source flux increases. This kind of correlation has been noted in X-ray binaries and Seyfert galaxies (Uttley \& M$^{\rm c}$Hardy 2001; Edelson \et 2002; Vaughan \et 2003a), and is due to a linear correlation between rms and flux (see also \S~\ref{sec:psdvar}). This effect therefore may not imply real non-stationarity of the variability. Panel 3 and 4 of  Figure~\ref{fig:lc:362} clearly demonstrates this:  the excess variances of 362-1 are systematically about one order of magnitude larger than those of 362-2, because exposure 362-1 was obtained in PN timing mode while 362-2 in PN imaging mode. In Figure~\ref{fig:lc:362}, the count rates of 362-1 were scaled down by a factor of about 5 in order to smoothly connect with 362-2. We can also demonstrate this effect by comparing excess variances of 362-1 with 174-1 and 450-3 (panel 3 and 4 of Figure~\ref{fig:lc:174}--\ref{fig:lc:450}). The three exposures show similar variability amplitude (about factor 2 from minimum to maximum count rate) but excess variances of 362-2 were about one order of  magnitude larger than those of 174-1 and 450-3.

Normalized variability amplitude (\fvar\ or \nxs) can ``filter'' such linear correlations. The bottom panels of Figures~\ref{fig:cor:174}--\ref{fig:cor:545} show the binned \fvar\ as a function of mean count rate, which was calculated in the same way as binned \xs . Interestingly, \fvar\ linearly anti-correlates with count rate for all observations, in contrast to the linear correlation between \xs\ and count rate. Therefore, the variability of PKS~2155--304 does show genuine (strong) non-stationarity, in the sense that the absolute rms (the excess variance) linearly increases with flux, but the fractional rms (the normalized excess variance or \fvar\ ), linearly decreases with flux. The mean \fvar\ is therefore inconsistent with a constant for both time and flux. However, given the small range of flux and variance changes, it is inadequate to determine and compare the slopes of the rms--flux relation and \fvar--flux relation for each observation. It is also important to point out that the rms--flux relations seen here are consistent with being linear, but the data are not adequate enough to rule out a power-law (nonlinearity) rms-flux relation (see Appendix D of Uttley, M$^{\rm c}$Hardy \& Vaughan 2005).

Given the implied steep PSD slopes of \pks, a red noise leak may be contaminating the variability amplitudes. The variability amplitude of a light curve that is a realization of a red-noise process may be dominated by trends on timescales much longer than the duration (500~s) considered here . This may especially be true for blazars during their flare states, where smooth increases or decreases in flux over tens of ksec may contribute greatly to a variability amplitude measured on time scales of 500~s and shorter. If the dominant flux trends associated with flares are steeper towards higher fluxes (the light curves of \pks seem flatter when the source is not flaring, for instance), this could potentially artificially produce a positive rms-flux correlation. Variability trends on timescales longer than 500~s should be removed from each segment (500~s long) before measuring the variability amplitude. For each segment, we did a least square fit to find the best-fitting line to characterize the overall dominating increasing or decreasing trend, subtracted the trend from the flux points, and then measured again the excess variance and \fvar. This yielded values of the excess variance that were only about few percents smaller than the previous values (without ``detrending'') since the variability contributions from red-noise leak were removed. However, this ``detrending'' did not change the flux-rms relation and flux--\fvar relation. Therefore, red noise leak does not significantly affect the excess variance and \fvar\ calculated between 25~s and 500~s. Furthermore, the steep PSD slope indicates that the intrinsic variability of the source is expected to be quite low at high temporal frequencies. We thus measured an unbinned periodogram for each light curve with a resolution of 10~s. The results showed that the intrinsic variability of the source is indeed much greater than the white noise variability due to Poisson noise on temporal frequencies between 0.002~Hz (1/500~s) and 0.02~Hz ($1/(2 \times 25$~s)).

\subsection{Root mean squared spectra}\label{sec:rmsspec}
 
The dependence of the variability amplitude on photon energy can, in principle, reveal spectral variability of a source. This dependence is called the root mean squared (rms) spectrum because the variability amplitude is measured with the common fractional rms variability amplitude, \fvar\ (e.g., Edelson et al. 2002; Vaughan et al. 2003b). A rms spectrum quantifies how the variability changes with photon energy. If \fvar\ (or normalized excess variance) is found to change significantly from one energy band to another, the PSD (slope or amplitude) must be energy-dependent. The changes of the variability amplitude with energy have been noted before in Mrk~421 and PKS~2155--304: the sources are systematically more variable toward higher energy (e.g., Fossati et al. 2001; Zhang et al. 1999; 2002 with \sax\ data; and Edelson \et 2001; Sembay et al. 2002; Ravasio et al. 2004 with \xmm\ data). 

However, given the length and binsize of a light curve, \fvar\ measures the integrated variability amplitude (see \S~\ref{sec:psdvar}). As TeV blazars (AGN in general) have steep ``red noise'' PSDs (e.g., Kataoka \et 2001; Zhang \et 2002), \fvar\ will be dominated by variations on the longest timescales probed by that light curve (e.g., Markowitz \& Edelson 2001). In order to probe short timescale variability we use a related parameter, called point-to-point fractional rms variability amplitude (\fpp; Edelson \et 2002), 
\begin{equation}
F_{\rm pp} = { 1 \over  \bar{x}  } \sqrt{ { 1 \over 2(N-1) }
{ \sum_{i=1}^{N-1} ( x_{i+1} - x_i )^2 } - \overline{ \sigma^2} }
\end{equation}
\fpp\ measures the variations between adjacent points (the sum of the squared difference of count rates between adjacent points, subtracted by the effects of measurement errors and normalized by mean count rate). For white noise, \fpp\ and \fvar\ are identical. However, for red noise variability, \fvar\ will be larger than \fpp, because the variations will be larger on longer timescales. \fvar\ and \fpp\ spectra measure the energy dependence of the variability at long and short timescales, respectively, so the ratio of \fvar\ to \fpp\ as a function of energy can, in principle, reveal the dependence of the PSD slope on energy: the larger the ratio, the steeper the PSD slope. If the ratio is energy-independent, the PSD slope is then energy-independent.

We split the light curve extracted from each exposure into 12 different energy bands (0.2--0.3, 0.3--0.4, 0.4--0.5, 0.5--0.6, 0.6--0.7, 0.7--0.8, 0.8--0.9, 0.9--1, 1--1.3, 1.3--2, 2--4, and 4--10~keV). The light curves in each band are then binned over 1000~s. They are strictly simultaneous and continuous except for 174-2 (there is a long gap due to high particle background). For each light curve, we calculated \fvar\ and \fpp\ spectra, and their ratio spectra, respectively. The error of \fvar\ and \fpp\ was calculated using equation (2) of Edelson et al. (2002), and the error for the ratio was propagated from the errors on \fvar\ and \fpp . The errors on \fvar\ and \fpp\ should be considered conservative estimates of the true uncertainty since they strictly assume the light curves are drawn from independent Gaussian processes. 

\begin{figure}
 \epsscale{1.0}
\plottwo{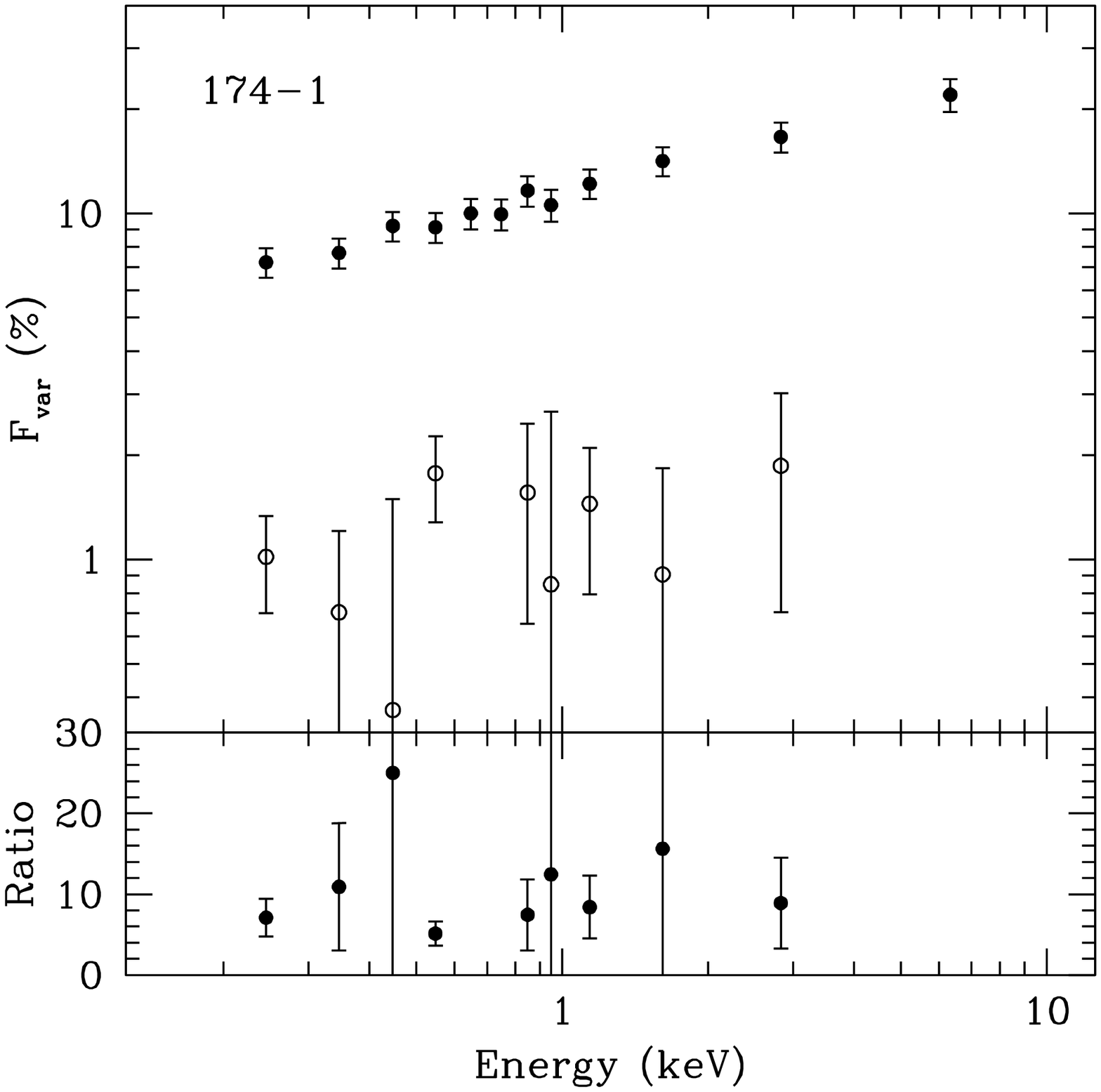}{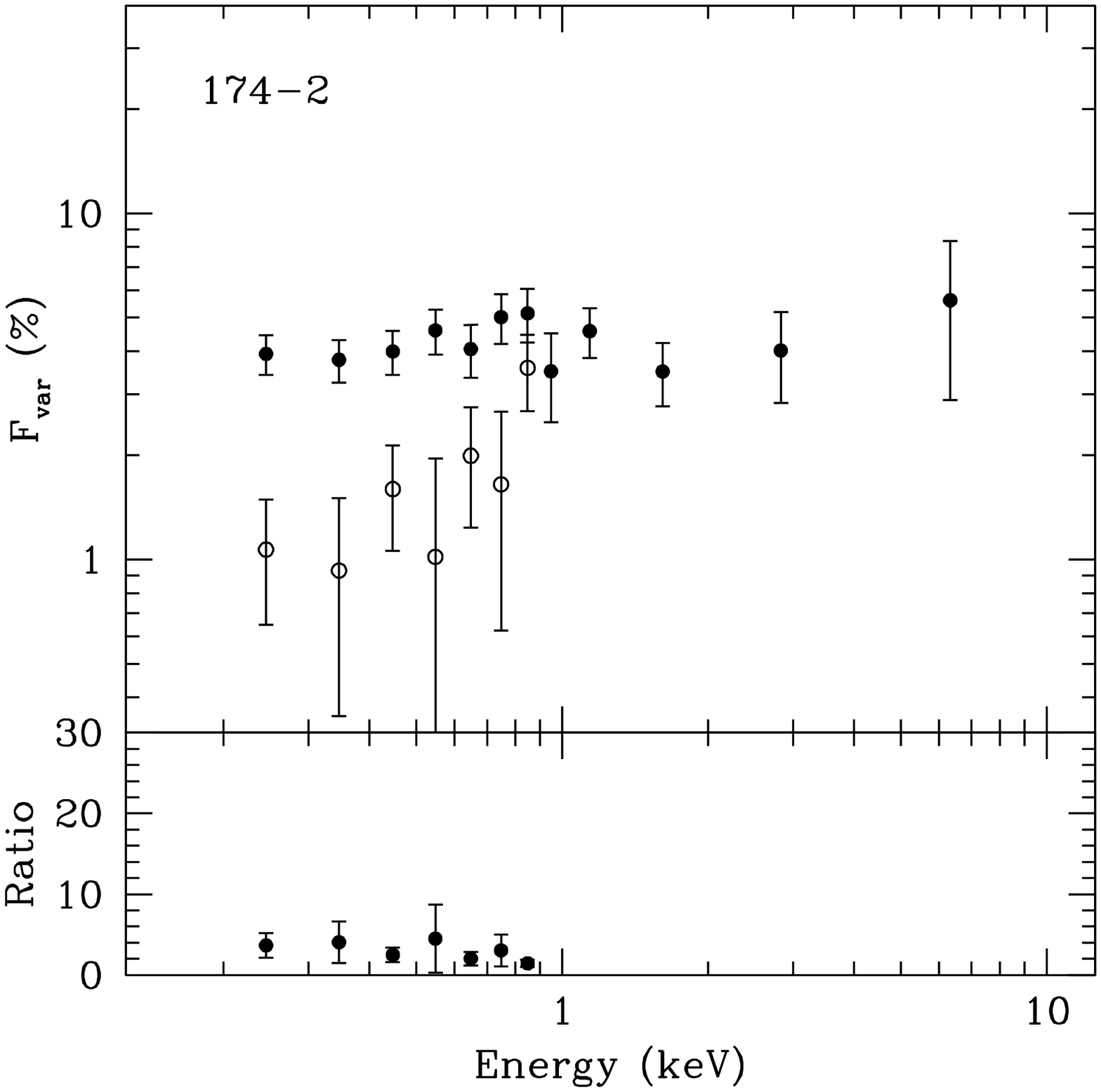}
\caption { \footnotesize Top panel: the fractional rms variability amplitude versus energy on two different timescales. The upper spectrum shows the variability amplitude integrated over timescales between the exposure length and 2000~s. Because of red noise variability, this spectrum is dominated by variations on timescale comparable to the length of the exposure. The lower spectrum represents the rms deviation between neighboring points (i.e., point-to-point rms), sampling variability only on short timescale ($\sim 1000$~s). Bottom panel: the ratio of the two rms spectra (i.e., the ratio spectrum). The ratio spectrum shows the way that the energy dependence of the variability amplitude changes with variability timescales, qualitatively representing the energy dependence of PSD. The errors were derived as described in the text.}
\label{fig:spec:174}
\end{figure}

\begin{figure}
\plottwo{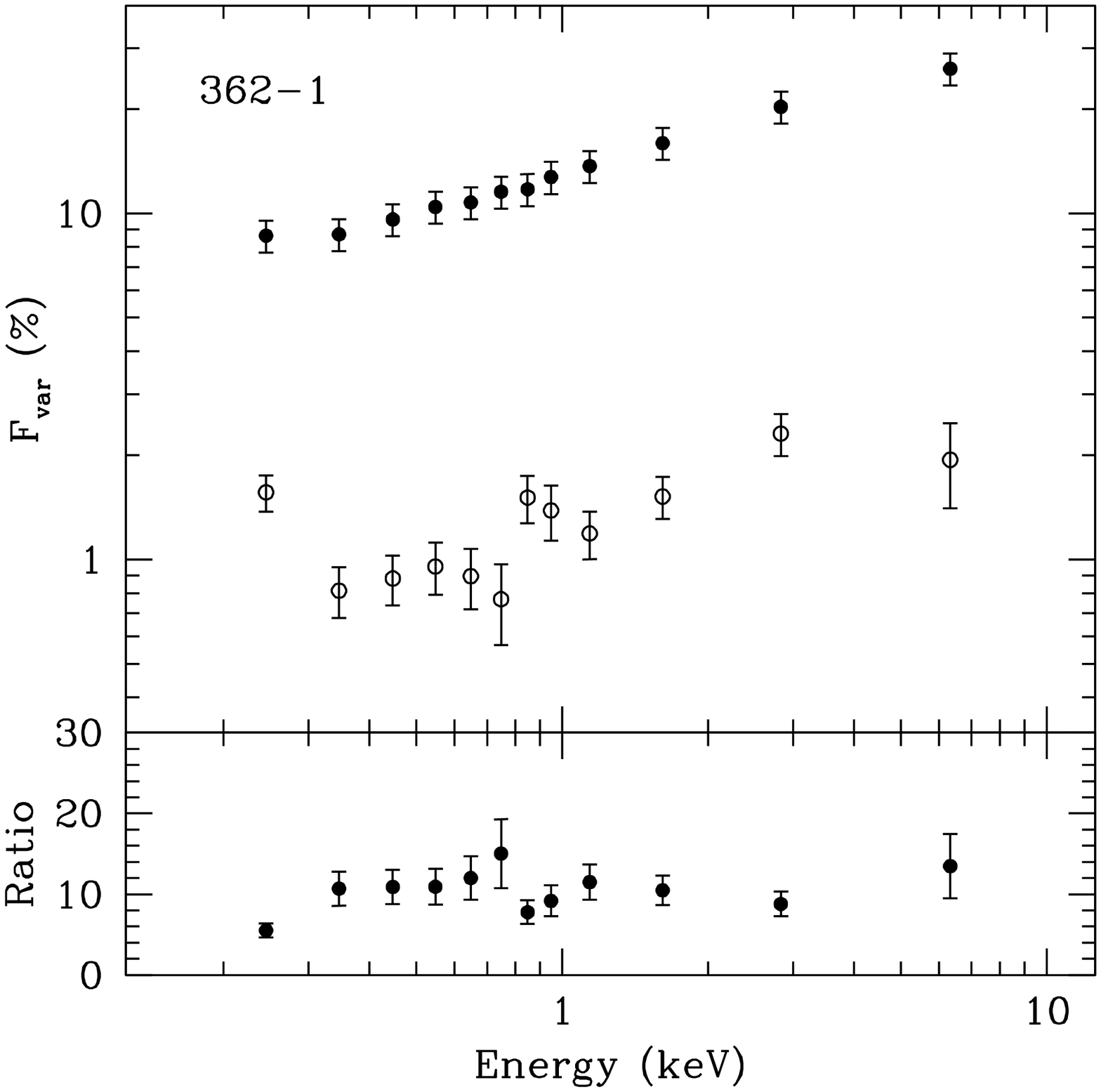}{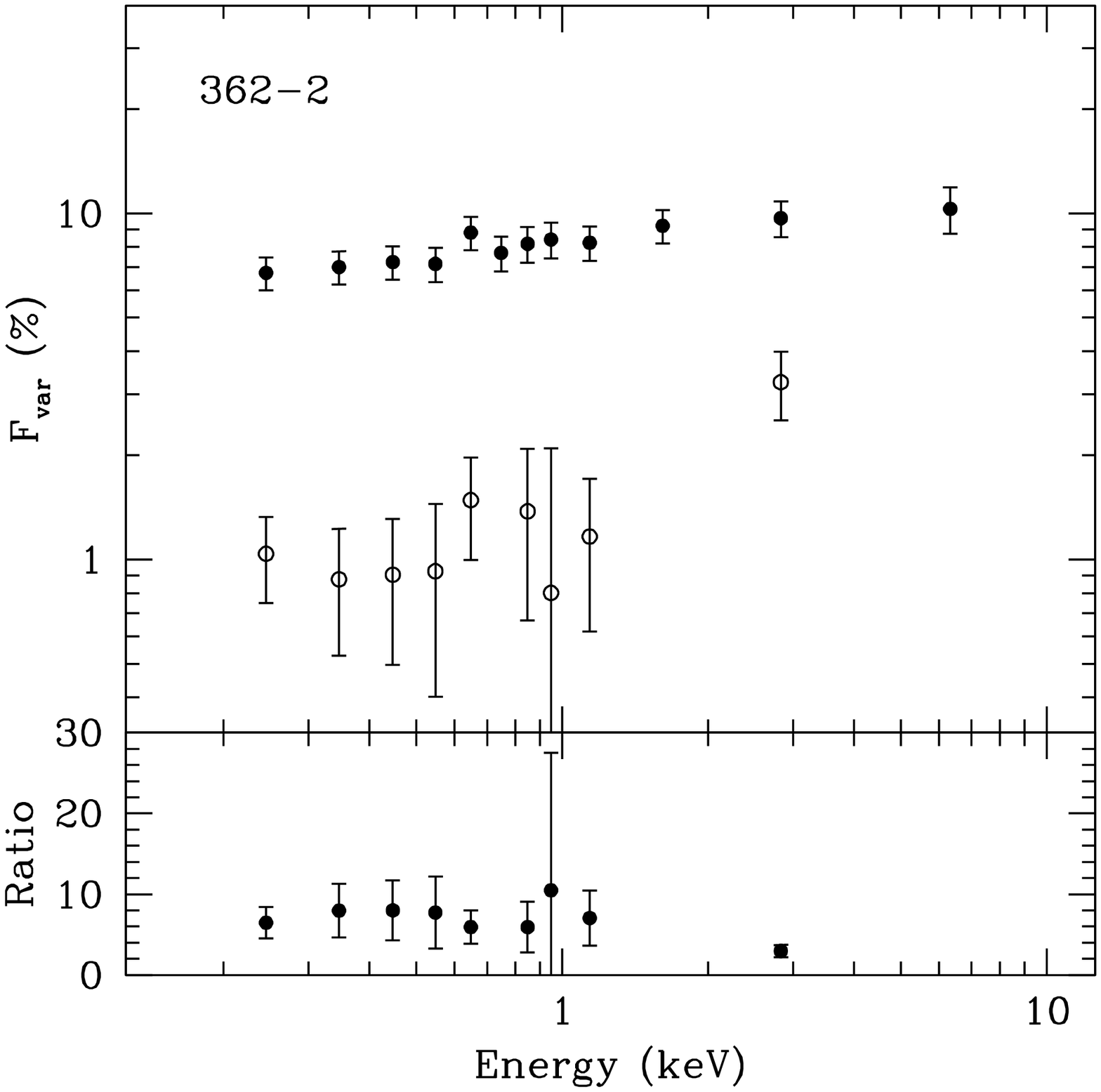}
\caption { \footnotesize Same as Figure~\ref{fig:spec:174} but using exposures 362-1 and 362-2 data.   }
\label{fig:spec:362}
\end{figure}

\begin{figure}
\plottwo{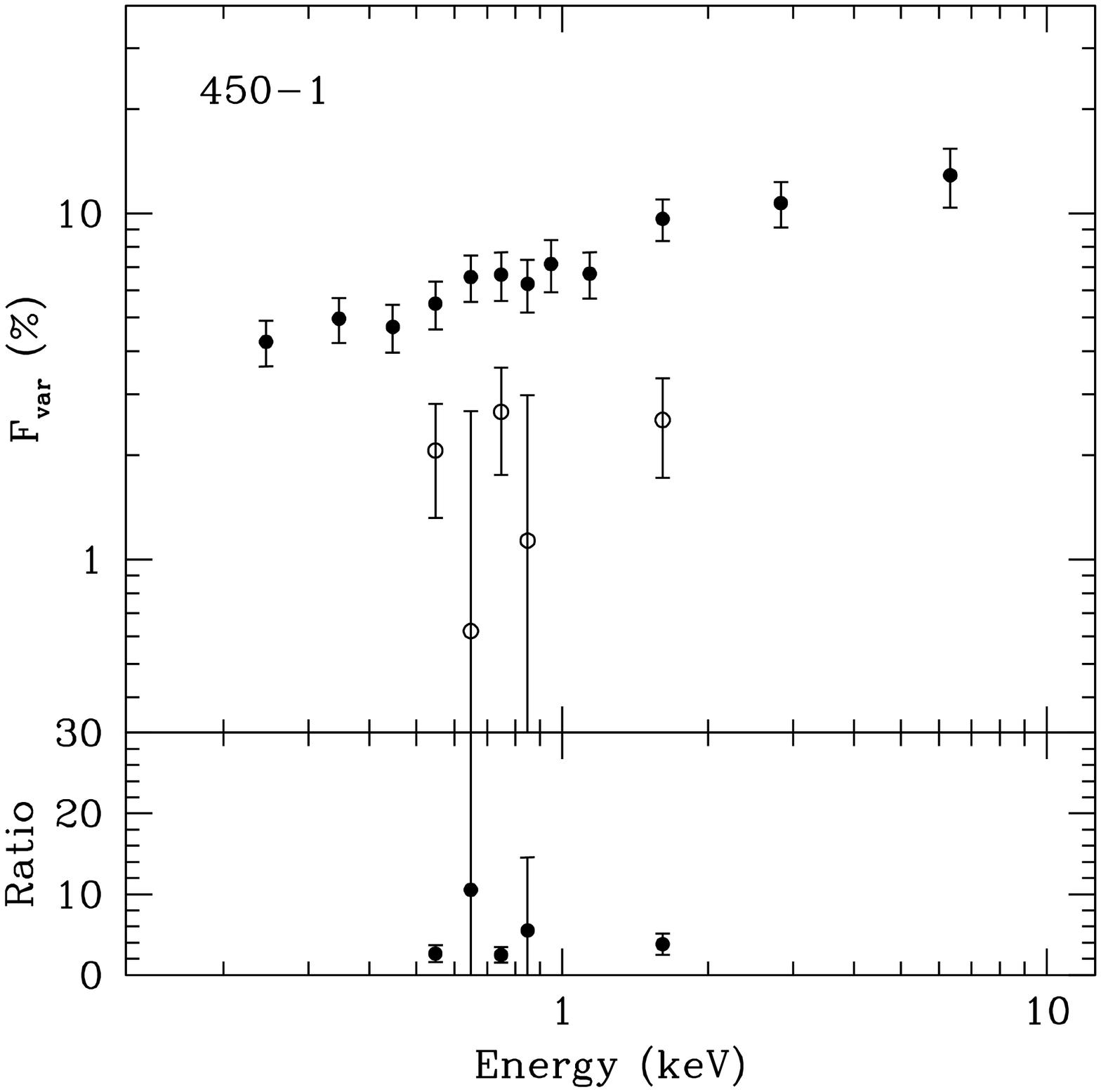}{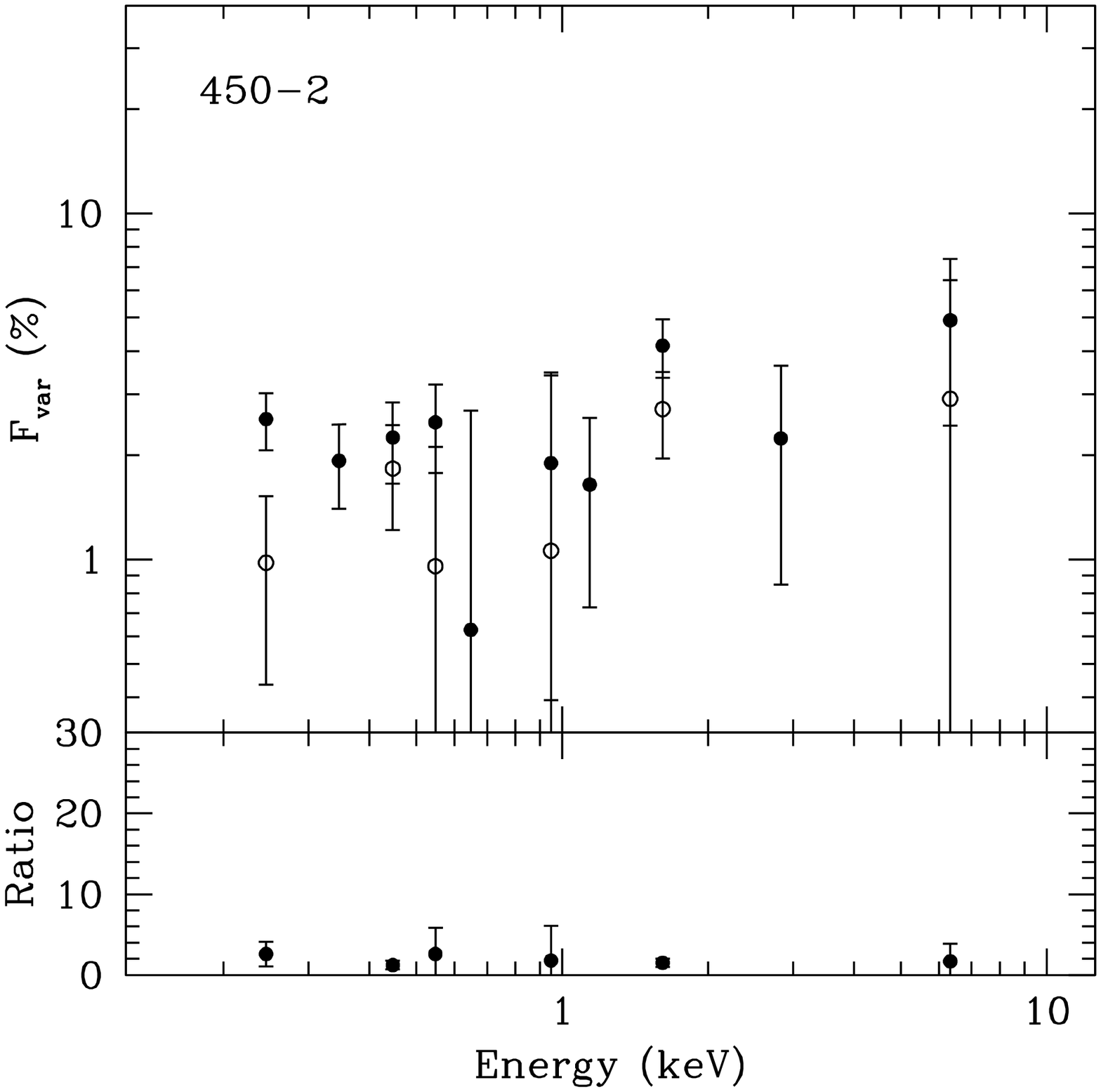}
\caption { \footnotesize Same as Figure~\ref{fig:spec:174} but using exposures 450-1 and 450-2 data.   }
\label{fig:spec:450}
\end{figure}

\begin{figure}
\plottwo{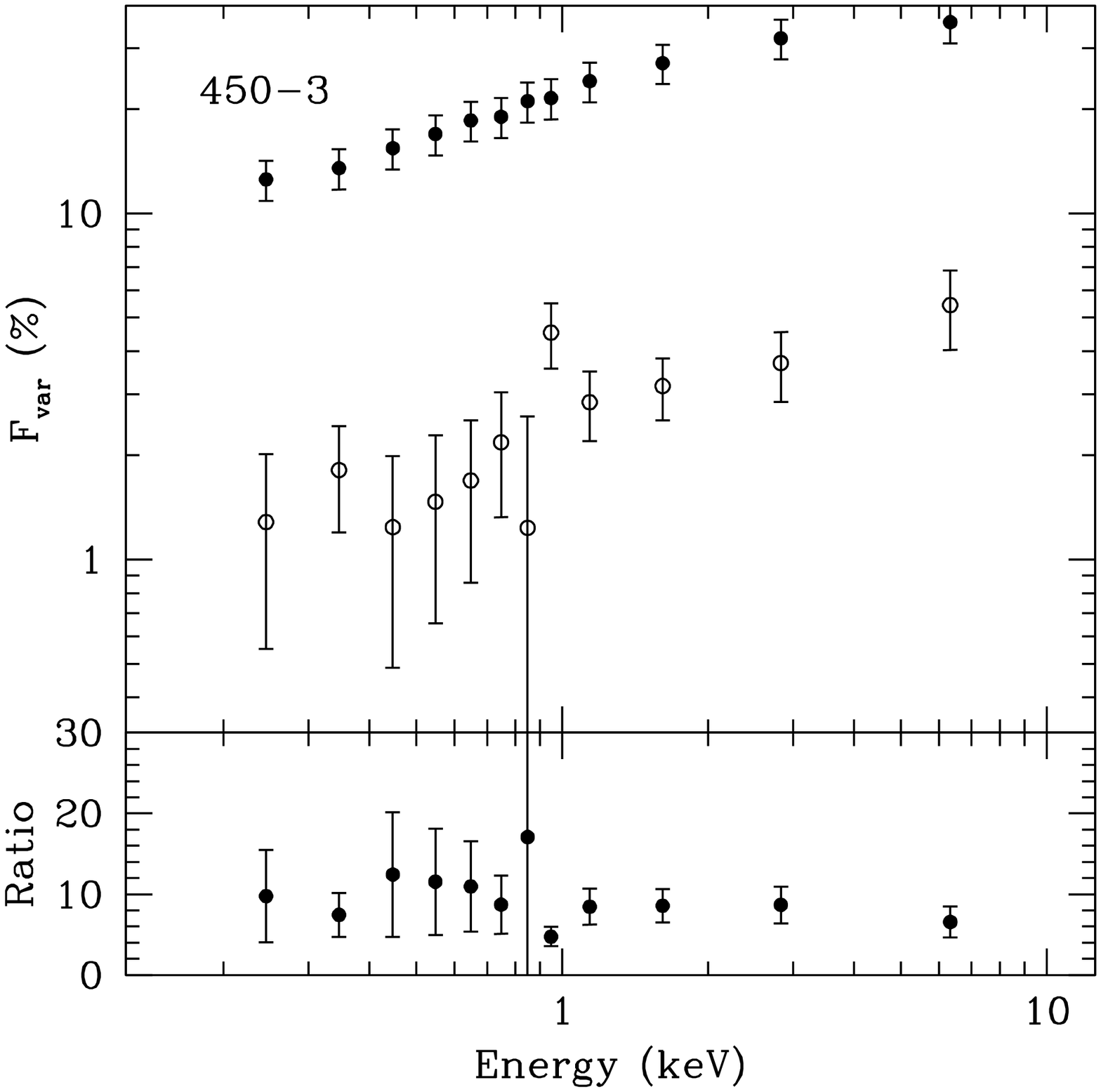}{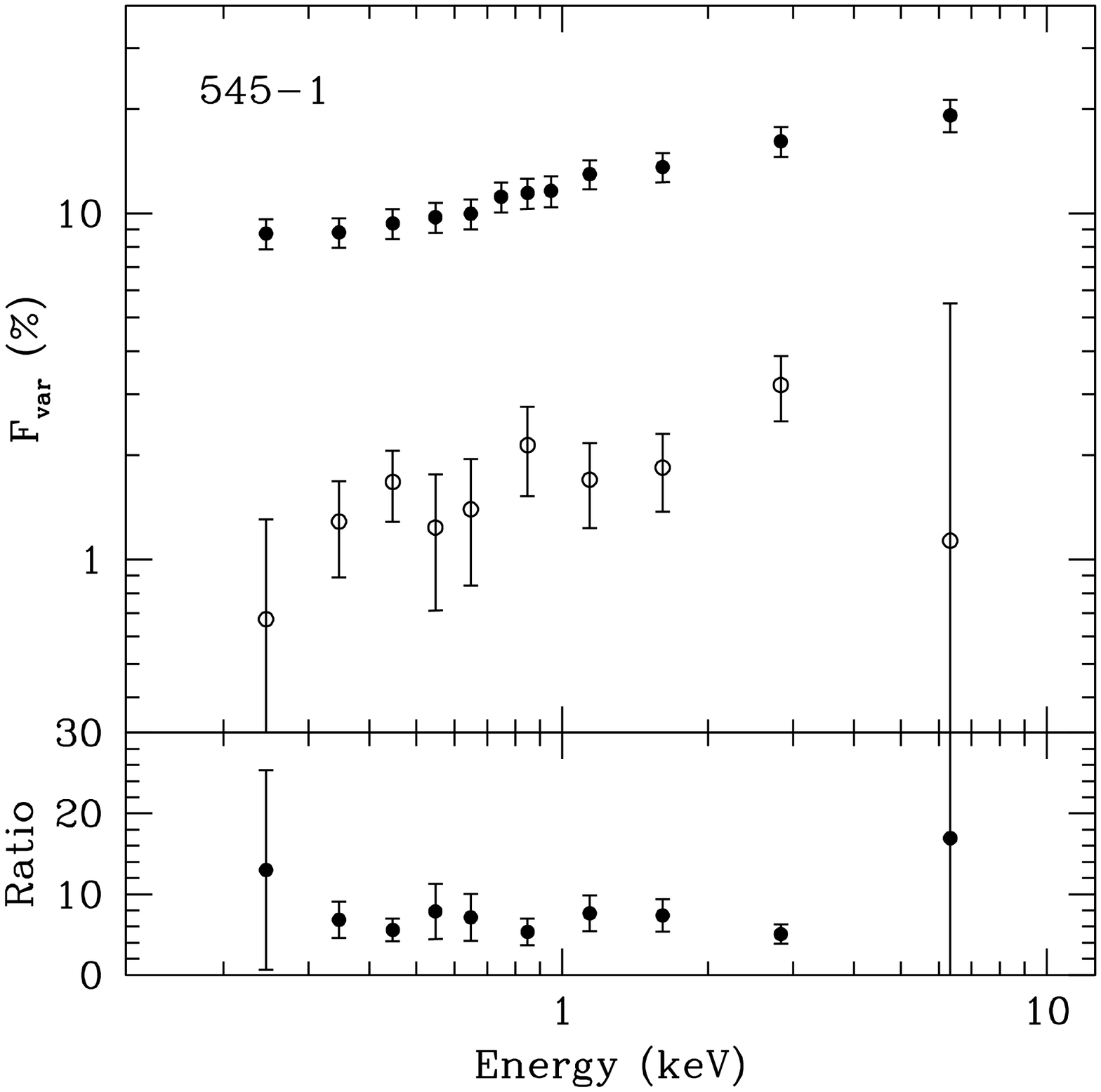}
\caption { \footnotesize Same as Figure~\ref{fig:spec:174} but using exposures 450-3 and 545-1 data.  }
\label{fig:spec:545}
\end{figure}

The upper spectrum of Figure~\ref{fig:spec:174}--\ref{fig:spec:545} shows the \fvar\ spectrum, the fractional rms variability amplitude integrated between the entire observation length and the Nyquist frequency (2000~s or $5\times 10^{-4}$~Hz), i.e., over about one order of Fourier frequency range. This frequency range  mostly has gaps determined by the periods of low-Earth orbital satellites such as \sax\ (Zhang et al. 2002). Since the variability is dominated by the longest timescale of a given exposure, the \fvar\ spectrum therefore reveals the energy dependence of the variations occurring on timescale comparable to the exposure length ($\sim 4\times 10^4$~s). It is clear from the figures that the \fvar\ spectra of PKS~2155--304 strongly depend on the energies during flare states (i.e., 174-1, 362-1, 450-3, 545-1), in the sense that the variability amplitude logarithmically increases with increasing energy. This trend indicates that the source experienced strong spectral variability during its active states. The \fvar\ spectrum of 450-1 also clearly depends on the energy but the variability amplitude is relatively small. The dependence of the variability amplitude on the energy is marginal for 362-2. The \fvar\ spectra of 174-2 and 450-2 were poorly determined and do not show clear energy dependence when the source was least variable. Moreover, it seems that the slope of such dependence differs from observation to observation. The results show that the variability of \pks\ strongly depends on the energy during flare periods while such dependence become marginal or even disappears when the source is less variable or in a quiet state. This phenomenology was previously found in the \sax\ data of Mrk~421 (Fossati et al. 2000). It is worth pointing out that the sampling of the \xmm\ light curves used here are consecutive, without gaps (except for 174-2). 

The lower spectrum of Figure~\ref{fig:spec:174}--\ref{fig:spec:545} shows the \fpp\ spectrum, measuring fluctuations between neighboring points and so probing the energy dependence of the variability on short timescales comparable to the bin-size ($\sim 1000$~s). However, the \fpp\ spectra were poorly measured due to small variations of the source on short timescales. In fact, in some cases the values of $F^2_{\rm pp}$ are negative due to large measurement errors, therefore these points are not present in the \fpp\ spectra of Figure~\ref{fig:spec:174}--\ref{fig:spec:545}. It is interesting to note that in some cases (e.g., 450-3) where the \fpp\ spectrum was relatively well determined, the \fpp\ spectrum tends to track the trend of the \fvar\ one, suggesting the ratio spectra of \fvar\ to \fpp\ is energy-independent. One can see from Figure~\ref{fig:spec:174}--\ref{fig:spec:545} that the ratio spectra do not show significant features, mostly keeping constant with energy. This further indicates that the PSD slope of \pks\ is possibly energy-independent but the PSD amplitude should be energy-dependent as the \fvar\ spectrum is energy-dependent, in the sense that the PSD amplitude is larger for higher energy. This phenomenology has been noted in the \xmm\ observations of Mrk~421 by directly comparing the PSDs in the soft and hard energy bands (see Figure 20 of Brinkmann et al. 2003).

\subsection{Estimate of Black Hole Mass}\label{sec:bhmass}

The variability amplitude of AGNs in terms of \nxs\ or PSD amplitude at a given timescale (or temporal frequency) anti-correlates with the source luminosity (Nandra \et\ 1997; Turner \et\ 1999; Leighly 1999; Markowitz \& Edelson 2001), and with the central black hole mass (\bhm) of AGNs (Bian \& Zhao 2003).  The long, well-sampled \xte\ light curves of a few Seyfert 1 galaxies revealed an unambiguous characteristic ``break frequency'', \nbf, at which the PSD changes in slope from 2 above to 1 below (Uttley \et 2002; Markowitz \et 2003). The break frequencies of Seyfert 1 galaxies are thought to be analogous to those of black hole X-ray binaries (BHXRBs) in a high state or the high-frequency break of BHXRBs in a low state. The break frequency, at which the PSD changes slope from 2 above to 1 below, appears to anti-correlate with \bhm\ from AGNs to BHXRBs, with the relationship $1/\nu_{\rm bf} \propto {\rm M_{BH}}$ (Uttley \et 2002; Markowitz \et 2003; Papadakis 2004). This is known as the linear scaling law of the PSD parameters with \bhm\ between AGNs and BHXRBs. However, the PSD break timescale -- $M_{BH}$ scaling relation of Markowitz \et (2003) \& M$^{\rm c}$Hardy et al. (2004) is consistent with being linear but the slope is in fact poorly constrained (if Cyg X-1 is not taken into account).

The break frequency more accurately represents the underlying dynamical process of the source. The relation $1/\nu_{\rm bf} \propto {\rm M_{BH}}$ can therefore be used to infer \bhm\ in AGNs by taking the \bhm\ of Cyg~X-1 as a reference point. This method however requires an unambiguous determination of \nbf\ in the AGNs, which was impossible before \xte. Instead, Hayashida \et (1998) and Czerny \et (2001) used the normalized PSD to estimate the \bhm\ for AGNs by calculating the ratio of the frequencies at which $\mathcal{P}(f)\times f$ has a certain (weell-defined) value for both the AGNs and Cyg~X-1. However, this method still requires rather long, high quality data that are available only for a few AGNs.

Equation~(\ref{eq:varpsd}) indicates that \nxs\ may also be used to estimate \bhm\ of AGNs. This led Nikolajuk, Papadakis \& Czerny (2004) to propose a simple method for estimating the \bhm\ for AGNs 
\begin{equation}
{\rm M}_{\rm BH} = C(T-2\Delta t)/\sigma^{2}_{\rm NXS},
\label{eq:bhm}
\end{equation}
where $T$ and $\Delta t$ are the light curve length and bin-size of a given observation. It is a very simple and straightforward method to estimate the \bhm\ as long  as the \nxs\ of the source has been estimated, and the constant $C$ is known. Given $T$ and $\Delta t$, Eq.~(\ref{eq:bhm}) indicates that the \bhm\ estimate is sensitive to the variability amplitude. The black hole mass of  Cyg~X-1 is known to be \bhm=10 M$_{\odot}$, so the constant $C$ can be estimated. By calculating the mean \nxs\ from a number of \xte\ observations of Cyg~X-1, Nikolajuk \et (2004) obtained $C=0.96\pm0.02$. When using Eq.~(\ref{eq:bhm}), several assumptions should be taken into account (Nikolajuk \et 2004): (1) the slope of the PSD is $2$ above \nbf; (2) \nbf\ scales linearly with \bhm; (3) the $\mathcal{P}(f)\times f$ amplitude is universal (Papadakis 2004); and (4) \nxs\ should be calculated from observations of length $T < 1/$\nbf . Moreover, the relativistic beaming effect of the emission from TeV blazars should be taken into account, Eq.~(\ref{eq:bhm}) is then modified as 
\begin{equation}
{\rm M}_{\rm BH} = C (T-2\Delta t)  \delta /\sigma^{2}_{\rm NXS},
\label{eq:bhm2}
\end{equation}
where $\delta$ is the Doppler factor. We take $\delta=18$ for \pks\ (Ghisellini \et 1998).  $\sigma^{2}_{\rm NXS}$ is the mean normalized excess variance averaged over a number of light curve segments.

\xmm\ observations of \pks\ produced several uniformly sampled light curves with high signal-to-noise (S/N) ratio. They are superior to the data obtained with previous satellites. We use the Rev. 174, 362, 450, 545-1 and 724 data to estimate the black hole mass of \pks. All 0.2--10~keV light curves were binned over 500~s. Since reliable PSD break frequency is not available for \pks\ (but see Kataoka \et 2001), for conservative purposes we broke each original light curve into several parts with each containing 20 points, thus the length of the light curves used is $T=10$~ks. We obtained a total of 33 such short but consecutive light curves. We calculated \nxs\ for each light curve. These \nxs\ estimates represent the variability amplitude integrated between 1~ks and 10~ks, almost corresponding to the gap in the PSDs derived with \asca, \sax, and \xte\ data (e.g., Zhang \et 2002). The mean of the 33 \nxs\ is $1.07 \times 10^{-3}$. Finally we used this mean \nxs\ to derive the black hole mass by using equation~(\ref{eq:bhm2}). The derived black hole mass is $\sim 1.45 \times 10^8 M_{\bigodot}$. Without correction for the relativistic beaming effect, the derived black hole mass would be $\sim 8.05 \times 10^6 M_{\bigodot}$, consistent with \bhm\ derived with the normalized PSD by Hayashida \et (1998) and Czerny \et (2001).

\section{Discussion}\label{sec:disc}

Variance estimates provide a simple but useful way of quantifying the variability of blazars. However, the stochastic nature of the red noise process means that only the average properties of the variance can provide physical insight.   Due to inadequate coverage, previous estimates of the variance in TeV blazars were based only on individual light curves. Because of the large scatter in the variance which is intrinsic to the red noise variability, the changes in individual variances usually provide less insight than expected. The important issue is how to disentangle the changes in the variance produced by the scatter in the variance of stationary process from real changes in the variance due to the underlying physical process responsible for the variability. In order to determine the variability of TeV blazars which exhibit the most significant X-ray variability among blazars, we have applied several methods, relevant to the variance estimate, to a number of high quality X-ray light curves of \pks\ obtained with \xmm.

We first calculated the individual variance estimates using 500~s long segments of the light curves binned on timescales of 25~s, and compared them with the expected scatter around the mean using the Monte Carlo simulation results of stationary processes (Vaughan \et 2003b). Our results showed that the observed individual variances of \pks\ were well within the confidence range expected for a stationary process. This is consistent with the idea that the underlying physical process responsible for the source variability is stationary. However, it is worth noting that the expected scatter in variance is rather large for the red noise data (more than one order of magnitude for a PSD with a slope of 2 adopted for \pks). The real changes in variability amplitude may be much smaller than this scatter. If this is the case, the real changes in variability amplitude can not be revealed by this method.

As averaging over many independent variance estimates reduces the random scatter inherent in the red noise data, real changes in the variability amplitude can be measured from the mean variances. There are two ways to average the variances. First, the averaged variance and its error can be calculated by binning the individual consecutive variance estimates, this gives rise to the evolution of the averaged variance with time. Whether the averaged variance changes with time or not can be tested through fitting the mean variances with a constant. Our results indicate the averaged variance changes with time during flare states, suggesting non-stationary variability. In the second way, the individual variance estimates were sorted and binned according to the count rates. In this way, the relationship between averaged variances and fluxes can be constructed. Interestingly, we obtained two kinds of correlations in the opposite senses: the mean absolute rms variability amplitude linearly correlates with flux, but \fvar\ linearly anti-correlates with flux. Although the first correlation may be artificial due to a linear correlation between variance and flux, the correlation between \fvar\ and flux should reflect real changes of the averaged variance with flux. Since random scatter in variance of the red noise variability should not give rise to any correlation between \fvar\ and flux, the anti-correlation between \fvar\ and flux shows that the underlying physical process responsible for the variability of \pks\ is not stationary. This genuine non-stationarity of the underlying process reflects real changers in the physical conditions of the variability process, such as physical parameters of the emitting region. 

The mean properties of a stochastic process have frequently been applied in the timing studies of X-ray binaries and Seyfert galaxies. The PSDs of X-ray binaries, estimated from the averaged periodogram of an ensemble light curves, evolves with time, indicating the variability process is strongly non-stationary (e.g, van der Klis 1995). It was also found in X-ray binaries and Seyfert 1 galaxies that the mean absolute rms variability amplitude scales linearly with flux of the source (Uttley \& M$^{\rm c}$Hardy 2001; Vaughan \et 2003b;), indicating the variability of these systems is intrinsically non-stationary.

More importantly, there may be two types of strong non-stationarity: (1) the strong non-stationarity due to the nonlinearity inherent in the rms-flux relation, and (2) the strong non-stationarity due to a changing PSD (which would yield averaged \fvar\ values that evolve with flux or time). In Seyfert galaxies, the rms-flux relation is one form of strong non-stationarity (since the averaged rms variability amplitude changes with time). However, this form of non-stationarity can be "factored out" by studying the \fvar--flux relation instead (Vaughan et al 2003b, Section 7.2). The \fvar-flux relation in Seyfert galaxies is flat (implying a PSD that is constant over those timescales), indicating no additional forms of strong non-stationarity. In blazars, the rms-flux relation also is a form of strong non-stationarity. However, even after removing this form of strong non-stationarity, the \fvar--flux relation is not flat, indicating the presence of some additional form of strong non-stationarity that is not present in Seyfert galaxies (e.g., implying a PSD whose shape and/or amplitude evolves with time or flux). This provides yet another suggestion that the X-ray variability mechanism in Seyfert galaxies and blazars may be different.

The differences between the variability amplitude of simultaneous light curves at different energy bands can be reliably examined using \fvar\ statistics. The rms spectra presented in this work were obtained on similar timescales but for different activity phases. The results show the shape of the rms spectrum is strongly energy-dependent during the flare phase, while during the quiet phase this dependence becomes marginal or even disappears. This indicates the emission from the source may be comprised of two components, one strongly variable and one almost steady. The variability of the variable component is strongly energy-dependent, but the non-variable one is not. Moreover, subtle changes in the shape of the rms spectra are also visible when comparing the individual rms spectra. These may be caused by parameter changes in the emitting region. The rms spectra of \pks\ are rather different from those of Seyfert galaxies (e.g., Vaughan \& Fabian 2004 for MCG-6-30-15; Vaughan \et 2003b for Mrk~766), indicating rather different emission mechanisms. The rms spectral shapes for MCG-6-30-15 also clearly changed with flux but in a different way than those for \pks: the rms spectra of \pks\ was monotonically increasing with energy while the rms spectra of MCG-6-30-15 are characterized by sporadic features (Vaughan \& Fabian 2004). This may be related to the fact that blazars are jet-dominated non-thermal systems while Seyfert galaxies are disk-dominated thermal systems. Furthermore, the point-to-point variability amplitude may track the rms spectrum, indicating the PSD slope of \pks\ is energy-independent, and in turn indicating the PSD amplitude should be larger with increasing energy as the rms spectrum is an increasing function of energy.

The black hole mass of some AGNs, including two blazars (3C~273 and \pks), has been inferred with the linear scaling law of the PSD characteristics between AGNs and Cyg~X-1. The PSD characteristics used includes the break frequency and a specific PSD value in $\mathcal{P}(f) \times f$ representation (e.g., Hayashida \et 1998; Czerny \et 2001; Uttley \et 2002; Markowitz \et 2003; Papadakis 2004). Hayashida \et (1998) derived the \bhm\ for \pks\ as $\sim 8.92 \times 10^6 M_{\bigodot}$. On the basis of the new PSD results for Cyg~X-1, Czerny \et (2001) found the \bhm\ of \pks\ as $\sim 2.45 \times 10^7 M_{\bigodot}$, a factor 2.8 larger. They determined the \bhm\ of \pks\ by assuming that the temporal frequency (or timescale) at $\mathcal{P}(f) \times f = 10^{-3}$ is linearly scaled with the \bhm, and taking \bhm$=10M_{\bigodot}$ of Cyg~X-1 as the reference point. However, calculating the PSD is very complicated, in particular, very few PSDs are available for blazars. The method proposed by Nikolajuk \et (2004) provides a simple way to infer \bhm\ in AGNs, which requires the estimates of \nxs\ only and should be applicable to many more cases than the methods based on the PSD. The assumptions of both methods are the same. Using this method, we obtained the \bhm\ of \pks\ is $\sim 8.05 \times 10^6 M_{\bigodot}$, consistent with the value obtained by Hayashida \et (1998). After correcting for the relativistic beaming effect, the black hole mass of \pks\ as $\sim 1.45 \times 10^8 M_{\bigodot}$ (assuming the Doppler factor $\delta=18$). Using \xte\ and \asca\ data, the \bhm\ of Seyfert galaxies estimated with \nxs\ are in consistent with those obtained with other methods (Nikolajuk \et 2004). 

 \xmm\ provides a large number of blazar light curves with a typical exposure time of $\sim 40$~ks. Owing to high output and long orbital period, the \xmm\  light curves are evenly sampled and have high S/N ratio. \xmm\ therefore provides us with a good opportunity to estimate the \bhm\ of blazars from \nxs, which can be compared with the \bhm\ obtained from other methods such as the stellar velocity dispersion from the optical spectra of the host galaxies of blazars (e.g., Barth, Lu, \& Sargent 2003; Falomo, Carangelo, \& Treves 2003). Since no values of the velocity dispersion of the galaxy \pks\ are available so far, its black hole mass have not yet been inferred with this (velocity dispersion) method. However, we note that for the dozen BL Lac objects where the velocity dispersion of the galaxy has been directly measured, one finds black hole masses of order $10^9 M_{\bigodot}$ for BL Lac objects (Falomo \et 2003). Moreover the observed host galaxy luminosity ($R=-24.4$, $H=-26.8$) also yields black hole mass of order $10^9 M_{\bigodot}$ for BL Lac objects (Kotilainen, Falomo \& Scarpa 1998; Falomo \et 2003). The black hole mass of \pks\ derived using X-ray variability is smaller by one order of magnitude than these, but the uncertainty is substantial. This possible inconsistency is worth studying with more X-ray data for more blazars, especially for blazars where the black hole masses have been estimated using the velocity dispersion of the host galaxy.

Finally, we mention caveats associated with using the normalized excess variance--black hole mass relation proposed by Nikolajuk \et (2004): (1) to do a direct comparison of variability amplitudes (or PSD shapes) between different classes of objects (e.g., blazars vs Seyfert galaxies/BHXRBs) would ideally require that the nature of the X-ray variability is the same in both classes of objects, but the link between the variability of jet-dominated X-ray emission and the variability of accretion disk corona-dominated X-ray emission is not clear. The fact that blazars may have steeper red-noise PSD compared to Seyfert galaxies could be viewed as an obstacle to "unifying" their X-ray variability properties. Another such obstacle is the fact that \fvar\ tends to increase as energy increases in (jet-dominated) blazars, while the opposite is generally true for (disk-dominated) radio-quiet Seyfert galaxies (e.g., above 2 keV; Vaughan \et 2003b for Mrk~766 and Vaughan \& Fabian 2004 for MCG-6-30-15); (2) the observation that \fvar\ decreases as flux increases in \pks\ could imply that the time-averaged PSD shape and/or amplitude changes as source luminosity increases. Physically, since the black hole mass of the source cannot evolve rapidly over the course of $\sim$~hours, this could mean that \fvar\ and the PSD of \pks\ may depend more on some other physical properties and less directly on black hole mass. In contrast, all Seyfert galaxies have a universal PSD shape which shifts towards lower temporal frequencies as one considers higher source black hole mass (and the variability is "slower" for higher-mass objects). However, PSDs for individual Seyfert galaxies do not seem to evolve on timescales of shorter than years (e.g., Markowitz \& Edelson 2001; 2004); (3) the different \fvar-flux relations indicate that the X-ray variability mechanism in Seyfert galaxies and blazars may be different. 


\section{Conclusions}\label{sec:conc}

Using simple variance statistics we have studied the X-ray variability properties of a TeV blazar \pks\ with several high S/N ratio and consecutive light curves obtained with \xmm\ EPIC-PN during a period of about 3 years. The main conclusions are as follows (even though comparison to other blazars should ideally take precedence):

\begin{enumerate}

\item The individual excess variance (or normalized variance) estimates, calculated with 500~s long segments of the 25~s binned light curves, are within the expected scatter around the mean for a red noise process, consistent with the assumption that the X-ray variability of \pks\ is the result of a stationary process. 
\item The mean excess variance and fractional rms variability amplitude, calculated by binning consecutive individual estimates in time, changes with time during flare states, indicating that the processes responsible for the X-ray variability of \pks\ is strongly non-stationary.
\item  However the mean absolute and fractional rms variability amplitude, calculated by sorting and binning the individual estimates according to count rates, shows an opposite correlation with flux. The absolute rms variability amplitude linearly correlates with flux, but the fractional rms variability amplitude linearly anti-correlates with flux. These two kind of correlations suggest that there may be two types of strong non-stationarity in the X-ray variability of \pks. The latter correlation supports strong non-stationary origin of the X-ray variability. 
\item During significant variations, \fvar\ shows a strong dependence on energy: the fractional rms variability amplitude increases logarithmically with increasing energy. The ratio of \fvar\ to \fpp\ is possibly energy-independent, suggesting the PSD slope of the source is energy-independent while the PSD amplitude should be larger for higher energy.
\item Using the normalized excess variance, we estimated the black hole mass of \pks\ as $\sim 1.45 \times 10^8 M_{\bigodot}$.

 \end{enumerate}


\acknowledgments
We are grateful to the anonymous referee for the constructive suggestions and comments that greatly improved the manuscript. We are also grateful to the scientific editor, Susan M. Simkin, for improving English of the manuscript. This work is based on observations obtained with \xmm, an ESA science mission with instruments and contributions directly funded by ESA Member States and USA (NASA). This work is conducted under Project 10473006 supported by National Natural Science Foundation of China, and under Project sponsored by the Scientific Research Foundation for the Returned Overseas Chinese Scholars, State Education Ministry. This work is also financially supported in part by the Directional Research Project of the Chinese Academy of Science under project KJCX2-SW-T08.


\clearpage
\begin{deluxetable}{ccrccccccc}
\tabletypesize{\footnotesize}
\tablecolumns{10}
\tablewidth{0pc}
\tablecaption{Observational Journal of \pks\ with \xmm\ EPIC-PN}
\tablehead{
\colhead{Rev} &\colhead{Obs Id} &\colhead{Exp Id} &\colhead{Date} 
&\colhead{Mode} &\colhead{Filter} 
&\colhead{Durat} &\colhead{Exp}
&\colhead{Bkg} &\colhead{Pileup} \\
& & &\colhead{(UTC)} & & &\colhead{(10ks)} &\colhead{(10ks)} & &
}
\startdata
0087 &0124930101 &87-1  &2000-05-30T10:20:09-30T20:53:29 &SW &Medium &3.80 &2.66 &yes &yes \\
     &0124930201 &87-2  &2000-05-31T00:52:59-31T17:21:38 &SW &Medium &5.93 &4.16 &yes &yes \\
0174 &0080940101 &174-1 &2000-11-19T19:00:40-20T10:55:39 &SW &Thin1  &5.73 &4.02 &no  &yes \\
     &0080940301 &174-2 &2000-11-20T13:15:19-21T05:25:19 &SW &Thin1  &5.82 &4.08 &yes &yes \\
0362 &0124930301 &362-1 &2001-11-30T03:12:05-30T15:30:29 &TI &Medium &4.32 &4.27 &no  &no  \\
     &           &362-2 &2001-11-30T15:54:06-01T04:17:30 &SW &Medium &4.46 &3.13 &no  &yes \\
0450 &0124930501 &450-1 &2002-05-24T11:18:09-24T20:08:10 &SW &Medium &3.18 &2.23 &no  &yes \\
     &           &450-2 &2002-05-24T20:31:33-25T05:19:52 &SW &Thin1  &3.15 &2.21 &no  &yes \\
     &           &450-3 &2002-05-25T05:43:16-25T14:01:35 &SW &Thick  &2.99 &2.10 &no  &yes \\
0545 &0124930601 &545-1 &2002-11-29T23:32:52-30T15:20:17 &SW &Thick  &5.67 &3.98 &no  &no  \\
     &           &545-2 &2002-11-30T15:57:29-01T07:14:54 &TI &Thick  &5.50 &5.44 &yes &no  \\
0724 &0158960101 &724   &2003-11-23T00:52:28-23T08:16:59 &SW &Thick  &2.66 &1.87 &yes &no  \\
\enddata
\label{tab:obs}
\end{deluxetable}

\clearpage

\end{document}